\documentclass[epjST]{svjour}

\usepackage{graphicx}
\usepackage{bm}
\usepackage{color}

\newcommand{\bi}{\begin{itemize}}
\newcommand{\ei}{\end{itemize}}
\newcommand{\ff}[1]{{\bf #1}}

\def\bR{{\bf R}}
\def\br{{\bf r}}
\def\bk{{\bf k}}
\def\bq{{\bf q}}

\begin{document}

\title{Realistic many-body approaches to materials with strong nonlocal correlations}

\author{F. Lechermann, A. I. Lichtenstein and M. Potthoff}

\institute{I. Institut f\"ur Theoretische Physik, Universit\"at Hamburg, Jungiusstra{\ss}e 9, 20355 Hamburg, Germany}

\abstract{
Many of the fascinating and unconventional properties of several transition-metal compounds with partially filled $d$-shells are due to strong electronic correlations. While local correlations are in principle treated exactly within the frame of the dynamical mean-field theory, there are two major and interlinked routes for important further methodical advances: On the one hand, there is a strong need for methods being able to describe material-specific aspects, i.e., methods combining the DMFT with modern band-structure theory, and, on the other hand, nonlocal correlations beyond the mean-field paradigm must be accounted for. Referring to several concrete example systems, we argue why these two routes are worth pursuing and how they can be combined, we describe several related methodical developments and present respective results, and we discuss possible ways to overcome remaining obstacles.
}

\maketitle

%%%%%%%%%%%%%%%%%%%%%%%%%%%%%%%%%%%%%%%%%%%%%%%%%%%
\section{Introduction}
\label{intro}

The quantum theory of the electronic structure of solids is very much characterized by the dichotomy of spatially {\em local} and {\em nonlocal} concepts. 
Nonlocality first comes across when considering the Hamiltonian of an electron moving in a potential $V$ with the discrete translational symmetries of an infinite three-dimensional lattice:
Its eigenstates are highly {\em nonlocal} and extend over the entire system, and the spectrum of eigenenergies, the band structure, lives in reciprocal or $\ff k$-space.
While this band-structure problem cannot be solved analytically, there are nowadays extremely efficient and highly reliable computational tools to solve the single-electron Schr\"odinger equation for rather general lattice potentials 
\cite{MJW71,And75,BSST90,BDGW92}.

Thanks to density-functional theory (DFT) \cite{HK64,KS65,JG89}, the solution of the band problem is key to a quantitative theory with predictive power: 
According to this paradigm of electronic-structure theory established in the 1960s, essentially independent electrons move in an effective one-particle lattice potential $V$ which, on a static mean-field level, includes effects of Fermi statistics (``exchange'') and of ``correlations''.
The success of band theory is very much due to the fact that $V=V(\ff r)$ is {\em local} -- opposed to the difficult-to-handle nonlocal exchange term of Hartree-Fock theory, for example.

Essential for a realistic modeling of the electronic structure is the treatment of electronic correlations. 
This central notion refers to all physics beyond the concept of a single Slater determinant and beyond a simple construction of the ground state by filling electrons, according to the Aufbau principle, into the nonlocal one-particle eigenstates up to the Fermi energy.
Opposed to configuration-interaction schemes for atomic or molecular systems, an explicit treatment of correlations by means of the full $N$-electron wave function is not practicable since for solids one has to consider a macroscopically large number of electrons.
It is thus advisable to shift the main focus to a different fundamental quantity of interest: 
Most suitable in the context of the Hohenberg-Kohn theory is the {\em local} electron density $n(\ff r)$.
This allows for an average and static mean-field-like description of correlation effects by means of the local-density approximation (LDA) and extensions \cite{KS65,JG89}.

In the recent decades, however, more and more material classes have been discovered, have been designed or simply have moved in the focus of materials science, where the DFT-LDA approach breaks down.
In most cases this is related to correlation phenomena like spontaneous magnetic order \cite{DDN98}, correlation-driven metal-insulator transitions \cite{Mot90} or high-temperature superconductivity \cite{OM00}.
This comprises $d$-electron materials and $3d$ transition-metal compounds, for instance.
Typical examples are oxides, such as  the intensively discussed unconventional cuprate-based superconductors, or pnictides, in particular iron-pnictide superconductors, cobaltates, etc.\ but also Mott or charge-transfer insulators like transition-metal monoxides, or simply late $3d$ transition metals as the classical band ferromagnets Fe, Co, Ni. 
These ``strongly correlated electron systems'' require a more explicit treatment of correlations which is conventionally done with the help of Feynman diagrams, i.e., in the context of many-body perturbation theory \cite{AGD64}.
The central physical quantity here is the single-electron propagator or Green's function $G(\ff r, \ff r', t, t')$ which is spatially {\em nonlocal}.
The Green's function is closely related to photoemission spectroscopy \cite{Huf07} and describes the propagation of the additional hole in the occupied part of the band structure that is left in the final state; it thus describes a nonlocal process.
Note that temporal homogeneity implies that $G$ depends on the time difference only and, via Fourier transformation, can be written as a function of the excitation energy $\omega$.

In the 1990s an important discovery was made for many-electron models with {\em local} interactions, such as the famous Hubbard model \cite{Gut63,Hub63,Kan63}:
The summation of all {\em local} (renormalized skeleton) diagrams contributing to the electronic self-energy can be achieved in practice by solving an effective impurity problem, the parameters of which must be determined self-consistently. 
This idea constitutes the dynamical mean-field theory (DMFT) \cite{MV89,GK92a,GKKR96,KV04}, a powerful nonperturbative and internally consistent mean-field theory which exactly accounts for the {\em local} quantum fluctuations while {\em nonlocal} correlations are treated on average only.
A practical DMFT calculation must be based on an ``impurity solver''.
There are different highly sophisticated variants available, based on Krylov-space techniques, quantum Monte-Carlo simulations or renormalization-group techniques, for example \cite{CK94,BCP08,GML+11,GTV+14,LHGH14}.
The success of DMFT is largely due to the fact that the importance of nonlocal correlations diminishes with increasing spatial dimension and that this effect is already relevant for three dimensions.
In fact, the DMFT is the exact theory of the Hubbard model on an infinite-dimensional lattice.
DMFT is also a versatile theory which is easily adapted to different lattice-fermion models and which has been extended in various directions. 

Nonlocal correlation effects, i.e., correlations beyond the DMFT, are key to understanding and controlling the electronic properties of several important material classes. 
Still, the dynamical mean-field theory can serve as a starting point. 
The purpose of the present work is to discuss electronic-structure theory and materials science from this perspective. 
Referring to a couple of concrete and typical examples, we review some essential theoretical aspects and recent methodical advances in this field.

Section \ref{sec:loc} provides an introduction to the issue of local and nonlocal effects and what can be described by means of the DMFT. 
As an instructive example we discuss the subtle competition between the Kondo effect and the indirect magnetic exchange in a periodic Anderson model on the triangular lattice. 
Even on the pure model level, however, there are important effects of nonlocal correlations beyond DMFT, in particular the feedback of nonlocal two-particle correlations on the electronic self-energy. 
Several proposals have been made to access nonlocal effects using DMFT as a platform for further improvements.
Besides cluster extensions \cite{MJPH05} or further diagram resummations \cite{TKH07}, this comprises the idea of dual-fermion theory \cite{RKL08}, reviewed in section \ref{sec:nonloc}, and also includes the extension to the dual-boson technique \cite{RKL12}, which is in addition capable of handling nonlocal {\em interactions}. 

The combination of DMFT with band theory is a fascinating vision with the final goal of establishing a parameter-free computational approach, covering largely different material classes.
This ongoing development, described in section \ref{sec:dftdmft}, is highly ambitious as it requires to unify nonlocal concepts from band theory with the locality paradigm of DMFT. 
The discussion of V$_{2}$O$_{3}$ as a concrete example sheds some light on the current state of the art. 
Novel materials-design ideas are touched upon by discussing application to the prominent problem of
oxide heterostructures. 
In view of realistic nonlocal self-energy effects, we discuss in section \ref{sec:co} spin-polaron physics in Na$_x$CoO$_2$, based on applied dual-fermion theory.

%%%%%%%%%%%%%%%%%%%%%%%%%%%%%%%%%%%%%%%%%%%%%%%%%%%
\section{Local vs.\ nonlocal correlations}
\label{sec:loc}

Let us start our discussion with an application of the DMFT to a ``simple'' model system. 
Without going too much into the formalism and by concentrating on the physics, this should help to qualitatively understand the capabilities of the DMFT but also show what is beyond a dynamic mean-field approach.

One of the hallmarks of strong correlations among itinerant valence electrons is the formation of local magnetic moments \cite{Anderson61}. 
Local-moment formation is constitutive for much of the physics captured by the periodic Anderson model (PAM) which is the most simple model for heavy-fermion materials \cite{Fisk_rev,Tsunetsugu97_rev}.
Generically, the PAM describes a band of light conduction electrons (``$c$ electrons'') of bandwidth $W$ hybridizing, with local hybridization strength $V$, with a narrow band of ``$f$ electrons'' located at energy $\varepsilon_f$. 
Since the $f$ band is narrow, Coulomb correlations are important and are taken into account by an on-site Hubbard interaction $U$. 

In the local-moment regime of the PAM for strong $U$, charge fluctuations on the local $f$ orbitals are effectively suppressed.
The single-electron spectrum exhibits two Hubbard bands located around $\varepsilon_f$ and $\varepsilon_f + U$ which are well separated from the $c$ band located around the chemical potential $\mu$, i.e., $\varepsilon_f \ll \mu \ll \varepsilon_f + U$, and each $f$ orbital is exactly occupied by one electron.
In this limit, the low-energy physics is perturbatively captured by the more simple Kondo-lattice model \cite{Tsunetsugu97_rev,Capponi00}, where the $f$ degrees of freedom are represented by quantum spins with $S_{f}=1/2$. 
Mediated by a super-exchange mechanism \cite{SW66}, the localized magnetic moments on the $f$ orbitals couple to the $c$-electron spins via a local antiferromagnetic exchange of strength $J  = 8V^{2}/U$. 
This coupling $J$ lifts the macroscopic ground-state spin degeneracy and quenches the residual entropy. 

There are however, two different and competing mechanisms, RKKY coupling and Kondo screening, which generate the famous Doniach diagram \cite{Don77} and corresponding quantum phase transitions \cite{Lohneysen_rev}.
The {\em nonlocal} Ruderman-Kittel-Kasuya-Yosida (RKKY) \cite{RK54} indirect magnetic exchange is mediated via magnetic polarization of the conduction electrons; it dominates for weak $J$, and is characterized by the effective coupling strength $J_{\rm RKKY}(\ff q) = - J^{2} \chi_c(\ff q,  \omega = 0)$ where $\chi_c$ is the $c$-electron spin susceptibility.
The Kondo screening of the $f$ magnetic moments is a {\em local} correlation effect which is already captured in the impurity variants of the Kondo or Anderson model and dominates at stronger $J$. 
Below the characteristic Kondo scale \cite{Kondo64,Hewson} given by $T_{\rm K} \propto e^{-W/J}$, the $f$ moment forms a singlet with a mesoscopically extended cloud of $c$-electron moments.

%-----------------------------------------------------------------------------------------------------------------------------------------
\begin{figure}
\centerline{
\resizebox{0.98\textwidth}{!}{
\includegraphics{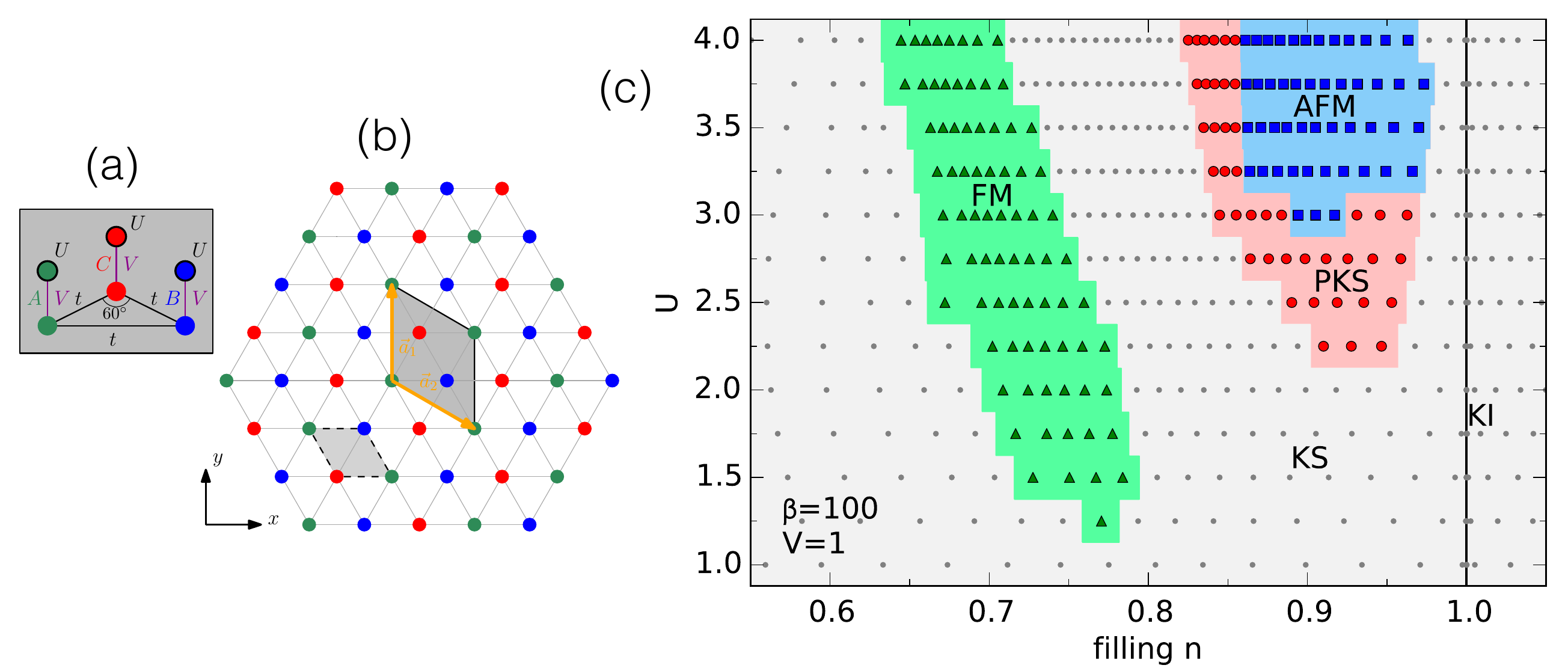}
}}
\caption{
Adapted from Ref.\ \cite{AAP15}: 
Magnetic phase diagram of the periodic Anderson model on the triangular lattice as obtained by site-dependent DMFT. 
(a) Unit cell used for the calculations. Sites A, B, C are treated independently.
For each site, an $f$ orbital with local interaction $U$ couples to a $c$ orbital via the hybridization $V$.
The nearest-neighbor hopping $t=1$ sets the energy scale.
(b) Triangular lattice with primitive cell (light gray, dashed lines) and the unit cell used for the calculations (gray, solid lines).
(c) Phase diagram in the plane spanned by $U$ and the filling $n$ (total electron number per site) at inverse temperature $\beta = 100$.
KI: Kondo insulator (at half-filling $n=1$), KS: metallic Kondo singlet state, PKS: partial Kondo-singlet phase, AFM:  antiferromagnetic phase, FM: ferromagnetic phase.
}
\label{fig:pam}
\end{figure}
%-----------------------------------------------------------------------------------------------------------------------------------------

Fig.\ \ref{fig:pam}c shows the phase diagram as obtained \cite{AAP15} by DMFT using the continuous-time quantum Monte-Carlo method (CT-QMC)~\cite{GML+11,RSL05} based on the hybridization expansion and the segment algorithm \cite{WCM+06} at a low temperature ($\beta = 1/k_{\rm B}T = 100$).
Each point corresponds to a converged DMFT calculation.
At half-filling $n=1$, the hybridization band gap in the noninteracting density of states results in a band insulator for $U = 0$. 
This develops into a correlated Kondo insulator (KI) with increasing $U$.
Furthermore, we find a metallic ferromagnetic phase (FM) at low fillings in the mixed-valence regime where the $f$ local moments are no longer well defined. 
Most interesting, however, is the competition between an antiferromagnetic phase (AFM) and the Kondo-singlet state (KS) in the local-moment regime for fillings slightly off half-filling.
As expected, the RKKY-induced AFM phase shows up for strong $U$, i.e., weak $J$, while the metallic heavy-fermion KS state, which connects to the KI at $n=1$, is realized at weaker $U$ but prior to charge fluctuations becoming dominant. 

The competition between Kondo screening and RKKY coupling is affected by the triangular lattice geometry (see Fig.\ \ref{fig:pam}b).
The nonbipartiteness of the lattice already explains the asymmetry of the phase diagram under particle-hole transformations: 
Choosing a positive hopping $t=1$ implies that the center of gravity of the noninteracting total density of states is located close to the lower band edge and that symmetry-broken magnetic phases are expected for fillings below half-filling. 

More importantly, however, the lattice also introduces frustration of the antiferromagnetic ordering tendency, and thus the problem gets more involved due to another energy scale which is associated with the release of frustration. 
For the PAM or the Kondo lattice, a mechanism of partial Kondo screening (PKS) \cite{MNY+10} has been suggested where a spontaneous, site-selective Kondo effect (say, on site A in the unit cell, see Fig.\ \ref{fig:pam}a) alleviates the frustration and thus allows the remnant spins (on sites B and C) to order magnetically via the RKKY coupling.  
This is an interesting compromise between a local and a nonlocal correlation effect which has attracted considerable attention in the past \cite{Ballou91,BFV11} and which has been studied in the PAM on the level of the static mean-field (Hartree-Fock) approximation \cite{HUM11,HUM12}.
The dynamical mean-field study discussed here correctly includes all local fluctuations and also captures the Kondo effect. 
Apparently this is sufficient to destroy any magnetic order at $n=1$ found in the static theory \cite{HUM11}. 

The phase diagram is actually obtained by applying a variant of DMFT (``site-dependent DMFT'') where the different correlated orbitals in the unit cell are treated independently, similar to a real-space DMFT approach \cite{PN97b}. 
In fact, a spontaneous breaking of the $120^{\circ}$ rotational symmetry is found at the border between the AFM and KS phase (Fig.\ \ref{fig:pam}c).
At this border it becomes favorable to avoid frustration by partial Kondo screening of one out of three $f$ moments. 
A detailed study \cite{AAP15} shows that the PKS state is metallic and that it supports an additional weak charge-density-wave ordering, mainly on the $c$ orbitals. 
Furthermore, due to proximity to the RKKY-coupled remnant moments, the corresponding breaking of time-reversal symmetry results in a slightly imperfect partial Kondo screening with a tiny residual magnetic moment on the $f$ orbital at the Kondo site in the unit cell.
DMFT predicts a robust PKS in a large parameter range.
Anisotropies \cite{MNY+10} are not necessary to stabilize the phase. 
PKS appears at noninteger fillings, i.e., the gain in kinetic energy might be essential to stabilize the phase such that spin-only models would be ruled out. 
Steering the system through the border between the paramagnetic heavy-fermion and the magnetically ordered phase could be an experimental route to detect the PKS phase.

There are some lessons learned about the dichotomy between local and nonlocal concepts in electronic-structure theory and about the point where DMFT stands:
The Kondo effect is due to a highly nonlocal singlet formation but is fully included in the framework of DMFT since it is driven by a local (exchange) interaction between a local magnetic moment and an uncorrelated bath.
The metallic Kondo singlet (KS) or heavy-fermion state of the PAM, on the other hand, is much more intricate as it must build from coherently overlapping Kondo singlets. 
DMFT can only provide an approximation to the physics on this coherence scale at energies even lower than the impurity Kondo scale. 
This influences the competion with the RKKY coupling in an essentially unknown way.
The RKKY interaction itself is a nonlocal interaction and thus cannot be treated explicitly with the standard single-site DMFT. 
In the PAM, however, it is generated perturbatively as an effective interaction. 
Hence, DMFT does capture its full spatial structure and therewith the corresponding tendencies towards magnetic ordering. 
However, DMFT does not include the feedback of nonlocal magnetic correlations on the one-particle self-energy, induced by the RKKY coupling, as only local diagrams are summed up.
Those missing fluctuations must result in mean-field artifacts. 
Typically, the DMFT will be biased, to some extent, toward magnetic ordering and a spontaneously symmetry-broken state $| \!\! \uparrow \rangle | \!\!\downarrow\rangle$ at the expense of a nonlocal singlet 
$(| \!\!\uparrow\rangle | \!\!\downarrow\rangle - | \!\!\downarrow\rangle | \!\!\uparrow\rangle)/\sqrt{2}$ \cite{TSRP12,ATP15}.
Although one could thus argue that the AFM state is overestimated in the phase diagram, compared to the Kondo singlet state, one would still expect a PKS phase at the border between both phases.
In any case, given the complexity of a problem posed by strong correlations in fermionic models on two-dimensional frustrated lattices with several mechanisms competing on low energy scales, the DMFT cannot be expected to provide more than a useful starting point. 
Cluster and other extensions of the single-site mean-field concept, see below, can be invoked to progressively include the effects of short- and long-range correlations. 

%%%%%%%%%%%%%%%%%%%%%%%%%%%%%%%%%%%%%%%%%%%%%%%%%%%
\section{Nonlocal correlations\label{sec:non}}
\label{sec:nonloc}

There are several strategies to include nonlocal correlations beyond the DMFT which can be explained by referring to the the Hubbard model. 
Here, we describe the dual-fermion \cite{RKL08} and the dual-boson approach \cite{RKL12} but start with a short discussion of DMFT again. 

Consider the noninteracting, ``kinetic'' part $H_{0}$ of the Hubbard model first. 
This is fixed by specifying the hopping-matrix elements $t_{ij}$ between sites $i$ and $j$. 
In the absence of the local Hubbard-interaction term, $H_{0}$ is easily diagonalized.
For a Hubbard model on a translationally invariant lattice with periodic boundary conditions, diagonalization is achieved by Fourier transformation to $\ff k$-space, and the one-particle excitations are fully captured by the ``band structure'' $\varepsilon(\ff k)$ with band width $W$. 
If, on the other hand, only the local part of the Hamiltonian is kept, i.e., the Hubbard interaction $H_{1}$ with interaction strength $U$ and the local term of $H_{0}$ fixed by the on-site  energy $\varepsilon_{0}$, the diagonalization of the Hamiltonian is trivial again and essentially reduces in real space to the diagonalization of a single ``Hubbard atom''. 

The great success of the DMFT is related to its ability to interpolate between these two extreme limits, or, at half-filling in particular, between the weak-coupling ($U/W  \ll 1$) metallic state and the strong-coupling ($U/W  \gg 1$) Mott-insulating paramagnetic state~\cite{GKKR96} in a nonperturbative way which treats the local correlations exactly.
On the operational level, DMFT replaces the correlated lattice problem (the Hubbard model) by an impurity problem with a single interacting site and with the same local Hubbard interaction strength $U$, typically by an Anderson impurity model. 
The one-particle parameters of this impurity model, which are fully determined by the local hybridization function $\Delta(\omega)$, must be determined self-consistently,
\begin{equation}
\Delta(\omega) = \omega - \varepsilon_{0} - \Sigma(\omega) - \frac{1}{G_{\rm loc}(\omega)} \, , 
\label{eq:dmftsc}
\end{equation}
from the (local) self-energy $\Sigma$ and the local element of the one-particle Green's function $G_{\rm loc}$ of the lattice model. 
The impurity problem, defined in this way, must be solved numerically to get the Green's function on the impurity $G_{\rm imp}$ and the local impurity self-energy. 
The self-consistency loop is closed by identifying the impurity self-energy with the lattice-model self-energy, assumed as local, and by calculating the lattice Green's function using Dyson's equation of the lattice model.

\begin{figure}[t]
\begin{center}
\includegraphics[width=0.45\textwidth]{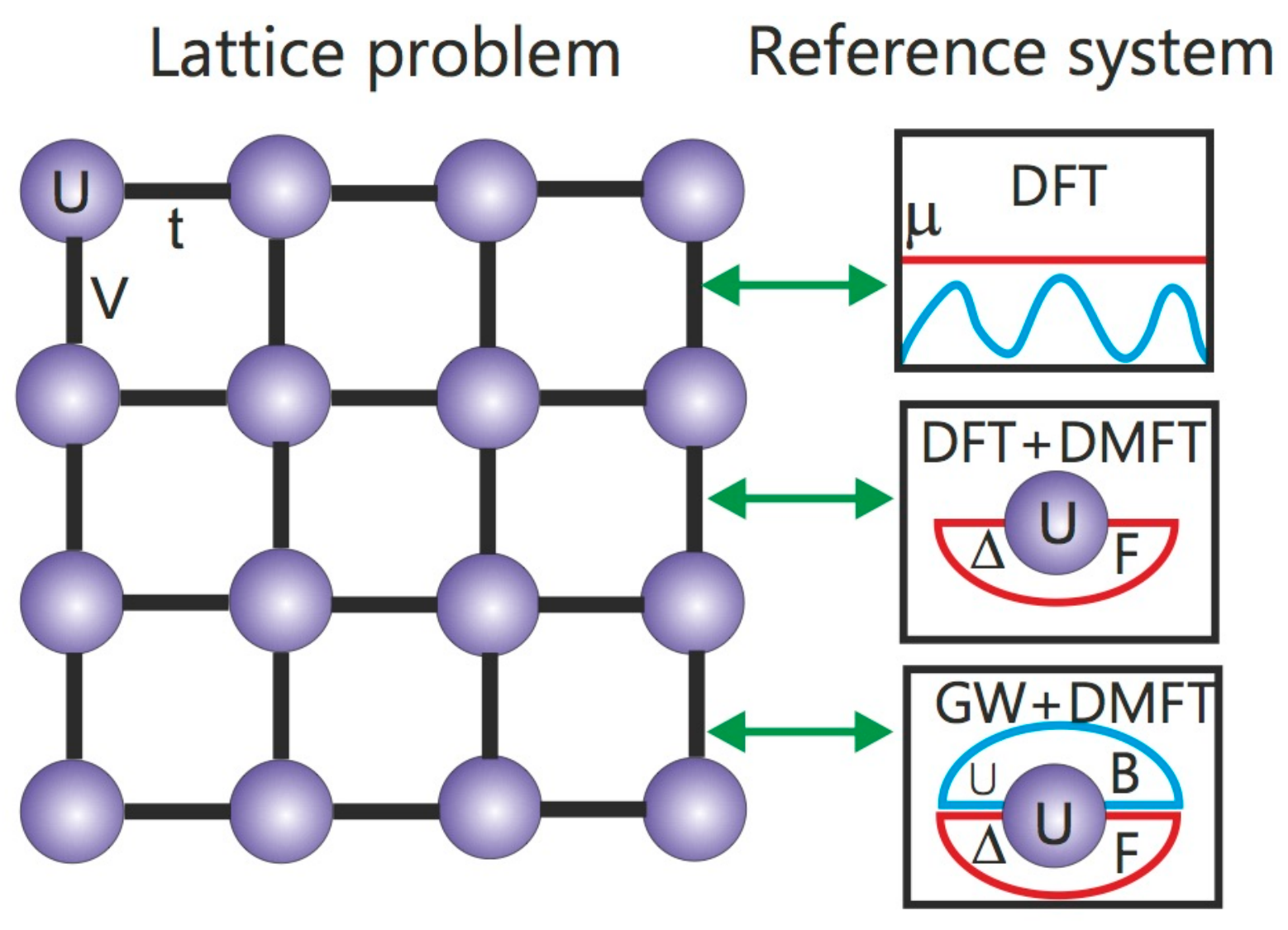}
\end{center}
\caption{
Sketch of three different approaches to describe many-body effects in lattice-fermion models: 
(i) Density-functional theory (DFT) with the interacting homogeneous electron gas as a reference system, defined by a constant external potential $\mu$.
(ii) Dynamical mean-field theory (DMFT) with an effective impurity problem as a reference system, i.e., a representative correlated atom embedded into a fermionic bath, specified by hybridization function $\Delta$.
(iii) GW+DMFT with a more complex impurity reference system, consisting of a correlated atom in a fermionic ($\Delta$) and a bosonic bath ($\Lambda$). The latter describes effects of the frequency-dependent screening of long-range Coulomb ($V$) interactions.
}
\label{fig:ref}
\end{figure}

A rather obvious idea suggests itself to incorporate nonlocal correlations beyond the DMFT: 
We start from a self-consistent DMFT solution defined by a self-consistent hybridization function $\Delta(\omega)$ of the Anderson impurity model. 
Since the Hubbard and the Anderson-impurity model share the same interaction part, one can think of the Hubbard model as the impurity model plus a residual term $\propto \varepsilon(\ff k) - \Delta(\omega)$ and treat this {\em perturbatively}.
This requires a novel perturbation theory. 
One may view this idea as a generalization of the Kohn-Sham idea in DFT of an optimal reference system, but with
a crucial difference. 
Here, not an interacting homogeneous electron gas, but an effective impurity model, tailored to the problem of strong correlations, serves as the reference system, see Fig.\ \ref{fig:ref}.

%-----------------------------------------------------------------------------------------------------------------------------------------
%-----------------------------------------------------------------------------------------------------------------------------------------

Since at the zeroth order of this perturbative expansion, i.e., on the level of the DMFT problem, we already have an interacting problem and since the perturbation is momentum and frequency dependent, one is forced to replace the Hamiltonians by actions within the path-integral formalism.
Note that the fermion path integral can also be used to formulate the DMFT itself \cite{GKKR96,kotliar_review}.
Now, the separation of the local and nonlocal terms is achieved by a Hubbard-Stratanovich transformation applied to the single-particle $(\varepsilon-\Delta)$-term. 
This provides us with a new action. 
Moreover, it is formally possible to integrate out the original local degrees of freedom and in this way generated an effective action in the transformed, so-called dual-fermion representation \cite{RKL08}. 
Note that integrating out the local degrees of freedom is not only a formal step but can be achieved in practice, namely by numerically solving the problem given by the local impurity action with the help of the CT-QMC method.

The dual action consists of a bare dual propagator $\tilde{G}_{0}(\ff k,\omega) = [G^{-1}_{\rm{imp}}(\omega)+\Delta(\omega) - \varepsilon(\ff k)]^{-1} - G_{\rm{imp}}(\omega)$, and a local but frequency-dependent effective potential related to scattering processes of two, three, and more dual particles on the impurity site.
The simplest two-particle dual potential coincides with the fully connected part of the impurity vertex 
$\Gamma_{\rm{imp}}^{\nu \omega \omega'}$, which can be calculated with the impurity CT-QMC solver
as a function of bosonic ($\nu$) and fermionic ($\omega, \omega'$) Matsubara frequencies. 
Normally, correlations between three particles on the DMFT impurity site are much weaker than two-particle correlations and can be ignored. 
The same applies to higher-order terms. 
Formalizing this argument, one can think of the dual-fermion formalism as an expansion in the order of local multi-particle correlation functions. 
This means that interactions between dual fermions are related with the connected part of the impurity vertex. Standard diagrammatic techniques can be applied for calculations of the full dual propagator $\tilde{G}_{0}(\ff k,\omega)$, which allows to obtain the nonlocal self-energy for the original fermions~\cite{RKL08} and to describe nonlocal correlations starting from the DMFT solution.

The dual-fermion approach is not necessarily bound to a specific starting point.
However, the DMFT starting point is very efficient. 
Namely, it corresponds to the elimination of all local diagrams for any $n$-particle correlation of dual fermions when using the DMFT self-consistency equation (\ref{eq:dmftsc}).
In the dual space, this simply reduces to $\sum_{\ff k} \tilde{G}_{0}(\ff k,\omega) = 0$ and means that, on average over the whole Brillouin zone, $\Delta(\omega)$ optimally approximates the electron spectrum $\varepsilon(\ff k)$, 
including its local correlation effects.
Therefore, the noninteracting dual fermions correspond to strongly correlated DMFT quasiparticles, and the remaining nonlocal effects can be quite small and reasonably described by, e.g., ladder summations of dual diagrams. 
This also explains the notion ``dual fermions''.

More interesting but also more complicated many-body nonlocal screening effects can appear if the original Hamiltonian contains additional nonlocal interactions $V(\ff q)$. 
In this case, one can formulate a similar so-called dual-boson theory \cite{RKL12}, which not only takes into account nonlocal fermionic propagators but also bosonic ones, which are screened by the long-range Coulomb interaction.
In this way, one can describe, e.g., plasmons or magnons in strongly correlated materials. 
The action of this ``extended'' Hubbard model reads as 
\begin{eqnarray}
S=-\sum_{\ff{k}\nu\sigma} [i\omega+\mu-\varepsilon_{\ff k}]c^+
_{\ff{k}\omega\sigma}c_{\ff{k}\omega\sigma}+\frac{1}{2}\sum_{{\ff q}\nu}
U_{\bf q}\rho^{*}_{\ff q\nu}\rho_{\ff q \nu} \; , 
\label{eq:action}
\end{eqnarray}
where the Grassmann variables $c^+_{\ff{k}\omega\sigma}$ ($c_{\ff{k}\omega\sigma}$) correspond to 
creation (annihilation) of an electron with momentum $\ff k$, spin projection $\sigma$, and fermionic Matsubara frequency $\omega$. 
The interaction $U_{\bf q}=U+V_{\bf q}$ consists of the on-site (Hubbard) term and the nonlocal long-range 
Coulomb interaction, respectively.
The screened Coulomb interaction can be frequency dependent as in the case of the constrained 
random-phase approximation (c-RPA), thus not producing any problems for the present formalism. 
The charge fluctuations are described by the complex bosonic variable $\rho_{\nu}=n_\nu-\langle n\rangle\delta_{\nu}$, 
where $n_\nu = \sum_{\omega\sigma}c^{+}_{\omega}c_{\nu+\omega}$ counts the number of electrons, and $\nu$ is a bosonic Matsubara frequency. 
For simplicity, we do not include spin degrees of freedom, which could be done by introducing vector spin-boson variables \cite{RKL12}.
Moreover we will consider only a one-band model, but keep the matrix form of all equations for a generalization to multi-orbital cases.

As discussed above, we split the lattice action, Eq.\ (\ref{eq:action}), into a sum of effective single-site impurity reference actions $S_{\rm imp}$, defined by the hybridization function $\Delta_{\omega}$ and by the screened local interaction $\cal U_{\nu}$, and into a remaining nonlocal part $\tilde{S}=\sum_{i} S^{(i)}_{\rm imp} + \Delta S$. 
The contributions to the latter are given by
\begin{eqnarray}
S_{\rm imp}=&-\sum_{\omega \sigma} [i\omega+\mu-\Delta_{\omega}]c^{+}_{\omega \sigma}c_{\omega \sigma} 
+\frac{1}{2}\,\sum_{\nu}\,{\cal U}_{\nu}\, \rho^{*}_{\nu} \rho_{\nu}
\; , \nonumber \\
\Delta S =&\sum_{\ff{k}\omega\sigma} \tilde{\varepsilon}_{\ff{k}\omega}\, 
c^{+}_{\mathbf{k}\omega\sigma}c_{\ff{k}\omega\sigma} +
\frac{1}{2}\,\sum_{{\ff q}\nu} \tilde{U}_{{\bf q}\nu} \, 
\rho^{*}_{{\ff q}\nu} \rho_{{\ff q}\nu} \; . 
\label{eq:rem_action}
\end{eqnarray}
Here, $\tilde{\varepsilon} _{\ff{k}\omega}=\varepsilon_{\ff k}-\Delta_{\omega}$ and $\tilde{U}_{{\ff q}\nu}=U_{\ff q}-{\cal U}_{\nu}$.
One can see that it is easy to incorporate the frequency dependence of the bare Coulomb interaction.
The strategy here is similar to the dual-fermion scheme and consists of an efficient perturbation scheme for $\Delta S$ in the action formalism. 
However, in addition to a fermionic Hubbard-Stratonovich transformation of the first term in Eq.~(\ref{eq:rem_action}), we also need to perform a standard bosonic transformation of the second (interaction) term. 
Following the same procedure as above and integrating out the original degrees of freedom, $c^+$ and $c$, with a proper rescaling of the fields, we arrive at the dual-boson action
\begin{eqnarray}
\tilde{S} &=
- \sum_{\ff{k}\omega} \tilde{G}_{0}^{-1}\tilde{c}^+_{\ff{k}\omega\sigma}
\tilde{c}_{\ff{k}\omega\sigma} - \frac{1}{2}\sum_{\ff{q}\nu} 
\tilde{W}_{0}^{-1}\tilde{\rho^{*}}_{\ff{q}\nu}\tilde{\rho}_{\ff{q}\nu}+ \tilde{V} \; ,
\end{eqnarray}
with the bare dual-fermion and dual-boson propagators, i.e.
$\tilde{G}_{0} = G_{\rm E} - G_{\rm{imp}} $ and 
$\tilde{W}_{0} = W_{\rm E} - {\cal W}_{\rm{imp}}$,
and with the dual interaction term $\tilde{V}$. 
We hereby also introduced the fermion ($G_{\rm E}$) and boson ($W_{\rm E}$) propagators for extended DMFT theory (E-DMFT)~\cite{kotliar_review}.
Note that, if we ignored interactions in the dual-boson action ($\tilde{V} =0$), our approach would exactly reduce to the E-DMFT, where the fermionic Green's functions depends on the local solution only ($G_{\rm{imp}}$) and the bosonic lattice propagator merely depends on the impurity screened interaction (${\cal W}_{\rm{imp}}$). 
In this case the E-DMFT self-consistent condition which connects the local part of the lattice propagators to the impurity ones gives solutions for the hybridization function $\Delta_{\omega}$ and the dynamical interactions ${\cal U}_{\nu}$.

Now the effective interaction in dual space not only consists of two-particle correlations (``square vertex'') but also contains effective electron-boson correlations (``triangular vertex'') which can be viewed as a result of partial bosonizations in the original diagrammatic series in powers of the interaction. 
In the standard dual-boson scheme one usually sums the ladder-type series of diagrams for the fermionic and bosonic propagators.

It is instructive to relate the dual-boson framework in its simplest approximation to the popular GW+DMFT 
scheme \cite{Biermann03,Sun02}, see Fig.\ \ref{fig:ref}.
In this approach, one neglects the two-fermion vertex and in the same approximation sets the effective fermion-boson vertex to unity. 
The fermions always interact with bosons in this scheme, which makes the E-DMFT approximation (or zeroth-order dual-boson quasiparticles) not as good as the DMFT one. 
We use the lowest-order correction in the electron-boson interactions which corresponds to the second-order boson self-energy:
$\tilde{\Pi}_{\ff{q}\nu }^{(2)} =-\sum_{\ff{k,}\omega }
\tilde{G}_{\ff{k}\omega }\tilde{G}_{\ff{k+q,}\omega +\nu }$.
In addition to the local DMFT self-energy we will have the non-local contribution from effective electron-boson
interactions. In the simplest RPA-like approximation this non-local part of GW+DMFT self-energy is equal to
\begin{equation}
\tilde{\Sigma }_{\ff{k}\omega }^{GW}=\sum_{\ff{q}\nu}\tilde{G}%
_{\ff{k-q,}\omega -\nu }\tilde{W}_{\ff{q}\nu }^{GW}\qquad,\quad
\tilde{W}_{\ff{q}\nu }^{GW}=\tilde{U}_{\ff{q}\nu }%
\left[ 1-\left( \chi _{\nu }+\widetilde{\Pi}_{\mathbf{q}\nu
}^{(2)}\right) \tilde{U}_{\ff{q}\nu}\right] ^{-1} \; .
\end{equation}
Let us point out two advantages of such a generalized GW+DMFT scheme: 
(a) the denominator of $\tilde{W}_{\ff{q}\nu}^{GW}$ contains the exact local susceptibility $\chi _{\omega }$, which improves the numerical accuracy of this approximation; 
(b) the lattice nonlocal interactions $U_{\ff{q}}$ needs to be corrected for the impurity local part $\mathcal{U}_{\nu}$ to give effective interactions $\tilde{U}_{\ff{q}\nu}$. 

%-----------------------------------------------------------------------------------------------------------------------------------------
\begin{figure}[t]
\begin{center}
\includegraphics[width=0.6\textwidth]{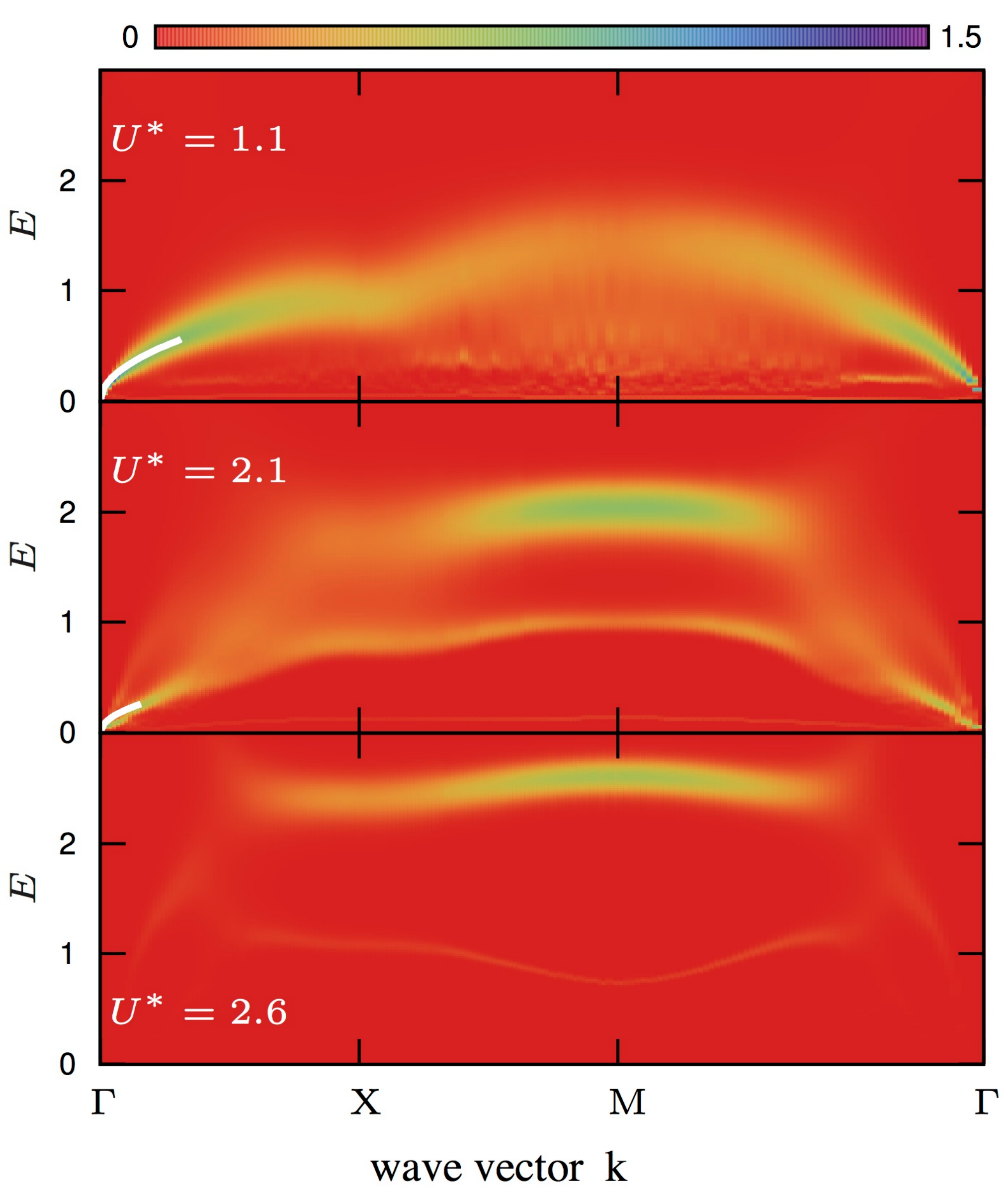} 
\end{center}
\caption{
Adapted from Ref.\ \cite{vanLoon2014}:
Inverse dielectric function as obtained by the dual-boson approach for the extended two-dimensional Hubbard model with long-range Coulomb interaction as a function of momentum and energy. Results are shown for three different values of the effective on-site interaction $U^{*}$.}
\label{fig:2dplasmons}
\end{figure}
%-----------------------------------------------------------------------------------------------------------------------------------------

Despite the recent successes of the $GW$+DMFT scheme, it does not provide a completely valid description of plasmons. 
This is due to an inconsistent treatment of the single- and two-particle properties, which breaks local charge conservation and gauge invariance. 
In case of a local and frequency-dependent self-energy in E-DMFT, vertex corrections from a local but frequency-dependent irreducible vertex are necessary to fulfill the Ward identity. 
In the dual-boson approach, this can be included via nonlocal polarization corrections, which are constructed diagrammatically. 
The resulting polarization vanishes in the long-wavelength limit at finite frequencies, as required by local-charge conservation \cite{Stepanov2016}. 
It therefore becomes possible to study the effect of strong correlations on the plasmon 
spectra \cite{vanLoon2014}, showing that the two-particle excitations exhibit both renormalization of the dispersion and spectral-weight transfer. 
This is similar to analogous interaction effects known from single-particle excitations.
Fig.\ \ref{fig:2dplasmons} shows the inverse dielectric function of two-dimensional surface plasmons in presence of long-range interaction $U_{\ff{q}}=U^{*}+V_{\ff{q}}$. 
For weak interaction one observes a broad particle-hole continuum and the expected $\sqrt{q}$ dependence of the plasmon dispersion at small $q$ in two dimensions. 
As the interaction increases, the plasmon dispersion $\omega_{p}^{2}(q)\approx\alpha V_q$ is renormalized. 
Two branches are clearly visible in the spectrum: 
The lower branch can be associated with particle-hole excitations between a Hubbard band and the quasiparticle peak, while the upper branch stems from excitations between Hubbard bands. 
Spectral weight is transferred from the lower to the upper branch as the interaction increases. 
Above a critical effective on-site interaction of $U^{*}\sim 2.4$, the system becomes a Mott insulator. 
In this state a two-particle excitation corresponds to the creation of a doublon and a holon. 
Such an excitation is highly localized, leading to a weakly dispersing band at energy $U^{*}$. 

%%%%%%%%%%%%%%%%%%%%%%%%%%%%%%%%%%%%%%%%%%%%%%%%%%%
\section{Realistic DMFT for correlated materials}
\label{sec:dftdmft}

The inclusion of nonlocal correlation effects using theoretical concepts as described, e.g., 
in the preceding section is an important route to go beyond the single-site DMFT. 
Another one is described in the following and may be termed ``realistic DMFT''. 
Its main idea is to combine the DMFT with density-functional theory (DFT), as will be discussed in 
section \ref{sec:ddbasics} below, with the overall goal of quantitative predictions for specific
materials rather than addressing model systems only. 
Already in seemingly standard-looking materials problems, multi-site and inhomogeneous aspects 
become relevant. Methodological aspects of the generalized real-space DFT+DMFT approach are 
mentioned in section \ref{sec:ddrealspace}, and the metal-to-insulator transition in V$_2$O$_3$ 
as a concrete example is discussed in section \ref{sec:v2o3}. 
A novel materials-design application of the given first-principles many-body theory is 
then discussed in section \ref{sec:het} for the context of oxide heterostructures.

\subsection{Combining DFT with DMFT
\label{sec:ddbasics}}

The promotion of DMFT to the realistic level by an adequate merging with electronic structure
theory in DFT has been envisaged already early on~\cite{GKKR96}. While the study 
of model Hamiltonians is a very important research field in order to investigate and reveal
key processes in correlated electron systems, concrete materials problems often harbor a vast
number of relevant characteristics that are hard to condense into a tailored (minimal) model problem 
from the start. Low-symmetry crystal structure, multi-orbital manifolds, intricate crystal-field
effects, structural distortions or sophisticated screening processes are only a few features that
may render an approach within a first-principles many-body scheme favorable~\cite{kotliar_review}.

Since DMFT builds up on a tight-binding perspective with well-localized orbitals to represent
interacting electron states, the linear-muffin-tin-orbital (LMTO)~\cite{And75} scheme was the 
first natural Kohn-Sham-based DFT framework to be allied with the many-body 
method~\cite{ani97,lic98}. Thereby, the LMTO orbitals (or parts of it) associated with selected 
states, e.g., a sub-manifold of the $3d$ 
shell of a transition-metal ion, serve as the so-called {\sl correlated subspace}. A different
combinational route, building up on the Korringa-Kohn-Rostocker method for the Kohn-Sham problem,
has a similar identification of the correlated states~\cite{min05}. In order
to utilize band-theory approaches which employ a plane-wave basis, the notion of local orbitals
has to be introduced. The spread of an inserted common LCAO basis is usually too large, but
Wannier-function methodologies, e.g., in the maximally-localized form~\cite{mar12},
are well suited. Thus using Wannier functions as a generic interface~\cite{lec06} allowed to open 
DFT+DMFT calculations to the widespread plane-wave community. The construction of explicit
special-featured Wannier functions for certain bands can sometimes be numerically cumbersome, 
especially for the case of low crystal symmetries and/or large supercells. An efficient 
alternative is the projected-local-orbital formalism~\cite{ama08,ani05} which 
uses a simple projection of Kohn-Sham states onto localized orbitals. This appears quite flexible 
and frees the correlated subspace from a direct band association. The correlated subspace should be 
quite generally be understood as a quantum-numbered real-space region where correlated electrons 
are most likely hosted. That space surely connects to the DFT charge density, but not necessarily 
needs to originate from a one-to-one mapping to a certain band dispersion. Once a correlated subspace
is selected, quantum impurity solvers based, e.g., on quantum Monte Carlo, Exact Diagonalization, etc.
yield the DMFT impurity solution.

%-----------------------------------------------------------------------------------------------------------------------------------------
\begin{figure}[t]
\centering
\includegraphics*[width=10cm]{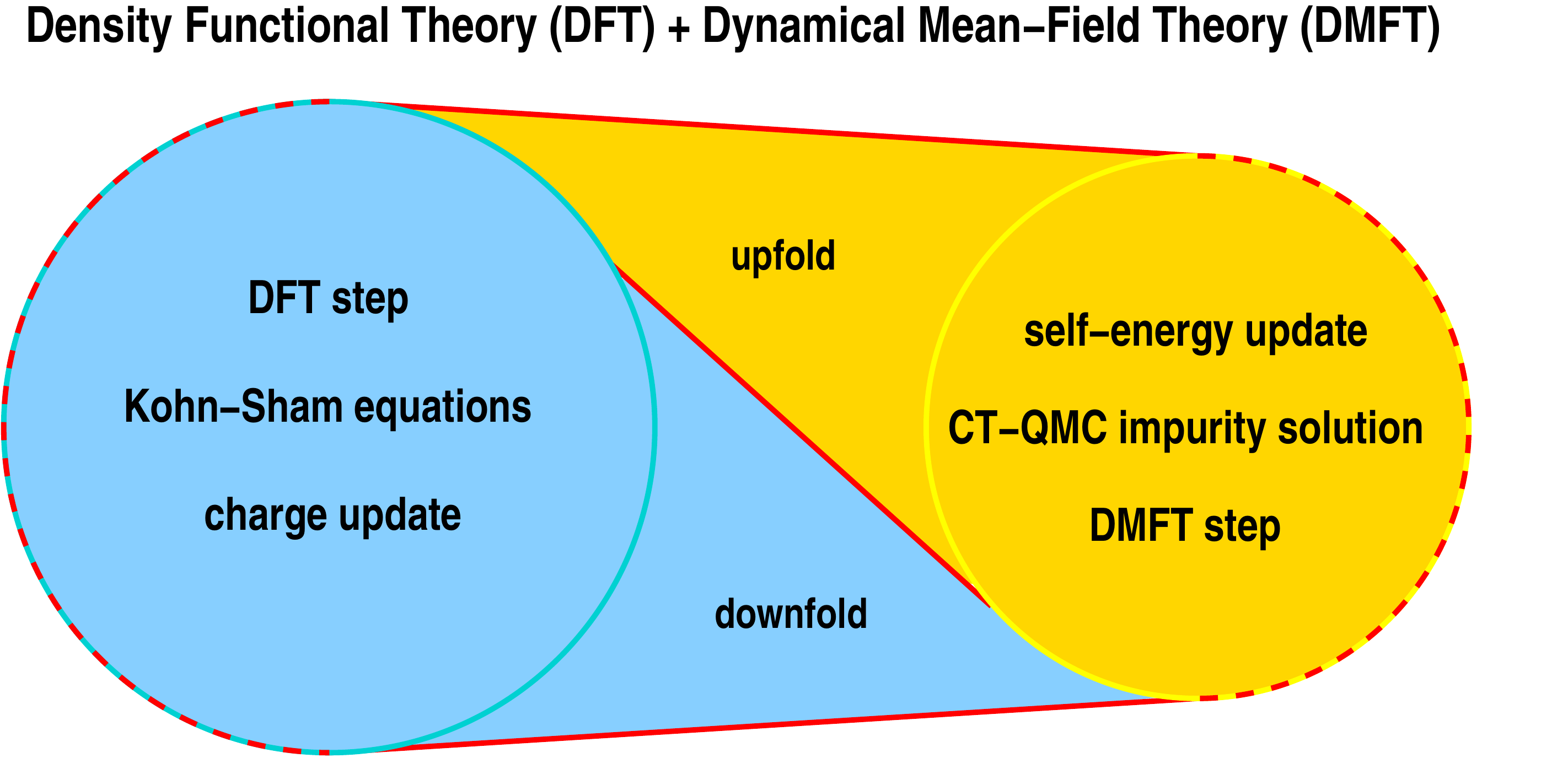}
\caption{Sketch of the charge self-consistent DFT+DMFT scheme pictured as a ratchet (for more
details see e.g.~\cite{lec06}). 
It starts with a DFT step by solving the Kohn-Sham equations with respect to the effective 
lattice potential $v_{\rm eff}$ and computing the resulting charge density. Downfolding to a 
correlated subspace defines a DMFT impurity problem and a solution step is undertaken, e.g., by 
using quantum Monte Carlo. The resulting self-energy is upfolded to the Bloch space and a novel 
$v_{\rm eff}$, based on the correlation-modified charge density, constructed. }
\label{fig:loop}
\end{figure}
%-----------------------------------------------------------------------------------------------------------------------------------------

So far we only dealt with the interface {\sl from} DFT {\sl to} DMFT, i.e., the full calculation 
employs the downfolding of a converged Kohn-Sham cycle via the named interfacing 
to a kinetic Hamiltonian that is then treated together with a suitable interacting Hamiltonian. 
This so-called ``one-shot'' or ``post-processing'' approach has been the state of the
art for several years. Its reliable to examine many spectral or local-moment 
properties~\cite{hel01,lic01,pav04,lec06}. However, since the realistic band-theory 
part and the interacting many-body part form the framework on equal footing, there has to
be in addition an interfacing {\sl from} DMFT {\sl to} DFT to close a self-consistency loop.
The charge self-consistent version of the method suits that goal. It computes an update of the,
now correlated, charge density subject to the many-body self-energies. Hence there is
an upfolding of the correlation effects to the complete Kohn-Sham (Bloch) Hilbert space. 
From the new charge density, a novel effective-single-particle potential for the next 
Kohn-Sham step is extracted and hence the full DFT+DMFT cycle~\cite{sav01,min05,pou07,gri12} 
is closed (see Fig.~\ref{fig:loop}). Using the normalized projections $\bar{P}$ between 
Bloch space and correlated subspace~\cite{ama08},
the key equations governing the down- and upfold interfacing read
\begin{eqnarray}
G^{\bR,{\rm imp}}_{mm'}(\omega)=&&
\sum_{\bk,(\nu\nu')\in {\cal W}}
\bar{P}^{\bR}_{m\nu}(\bk)\,G^{\rm bloch}_{\nu\nu'}(\bk,\omega)\,
\bar{P}^{\bR*}_{\nu' m'}(\bk)\quad,\label{eq:g_limband}\\
\Delta\Sigma^{\rm bloch}_{\nu\nu'}(\bk,\omega)=&&
\sum_{\bR,mm'}\bar{P}^{\bR*}_{\nu m}(\bk) \,\Delta\Sigma^{\rm imp}_{mm'}(\omega)
\,\bar{P}^{\bR}_{m'\nu'}(\bk)\quad.\label{eq:sig_limband}
\end{eqnarray}
Here, at self-consistency, $G^{\bR,{\rm imp}}_{mm'}$ is the impurity Green's function on
site $\bR$ with orbital indices $m,m'$ and $\Delta\Sigma^{\rm bloch}_{\nu\nu'}$ is the
$k$-dependent self-energy in Bloch space after double-counting correction with band indices
$\nu,\nu'$. As for the correlated subspace, there is a choice for the range ${\cal W}$ 
of included Kohn-Sham bands in the calculation.
The double-counting correction, similar as in static DFT+U~\cite{ani-u91,ani93,czy94}, 
takes care of the 
fact that some correlations are already treated on the DFT level. Note that albeit the latter 
equations are written in spirit of single-site DMFT, the Bloch self-energy is indeed naturally 
$k$-dependent because of the $k$-dependence of the projections $\bar{P}$. In other words, since 
a given orbital has varying contribution to given Kohn-Sham bands in reciprocal space, so has 
the associated self-energy. It is also instructive to reproduce the expression for the
correlated charge density $\rho$, i.e.
\begin{equation}
\rho(\br)=\sum \limits_{\bk,\nu\nu'}
\langle \br \vert \Psi_{\bk \nu} \rangle
\Bigl(f(\tilde{\epsilon}_{\bk \nu})\delta_{\nu \nu'}+
\Delta N_{\nu \nu'}(\bk)\Bigr)
\langle \Psi_{\bk \nu'} \vert \br \rangle\quad,
\label{eq:rho}
\end{equation}
where $\Psi$ denotes Kohn-Sham states, $f$ the associated
Fermi function and $\Delta N$ is the DMFT self-energy correction term~\cite{lec06,ama08}. 
Without the latter, the charge density reduces to the original Kohn-Sham DFT form (and also
realized in static DFT+U), resulting from contributions diagonal in the band index. However 
since a pure band picture is not vital in a many-body system and {\sl real-space} excitations 
also matter, additional terms off-diagonal in the band index contribute in the correlated 
regime. 

With this establishment of DFT+DMFT, correlated total energies become available and
permit the investigation of, e.g., phase stabilities~\cite{gri12}. Note, finally, that this
first-principles many-body scheme essentially works at finite temperature $T$. Electron states are
subject to the full thermal impact, resulting, e.g., in a well-defined description of paramagnetism
or the evaluation of coherence scales in electron transport.

This very brief sketch of the realistic-DMFT development described the matter on an informal level. 
Importantly, however, the approach may also be much more formally presented on the level 
of many-body functionals of Luttinger-Ward type. We refer 
to~\cite{geo04,kotliar_review} for detailed reviews on such formulations.

\subsection{Real-space DFT+DMFT
\label{sec:ddrealspace}}

The correlated subspace in many quantum materials invokes not only a single lattice site. 
For instance, multi-component compounds often harbor various symmetry-equivalent 
or -inequivalent correlated transition-metal sites. But this is easy to handle, 
since an impurity problem can be defined for each symmetry-inequivalent site, whereby the
coupling of those arises via the DMFT self-consistency condition~\cite{PN97b}. The self-energy 
of equivalent sites is determined for a representative site and transferred
to the remaining sites via the proper symmetry relations.

This multi-site or real-space generalization of single-site DMFT, of course, also neglects in
its realistic version explicit inter-site self-energy terms, i.e., explicit nonlocal correlations 
are absent. Yet especially fostered in the charge self-consistent framework, there is surely 
an implicit coupling, sufficient to sustain certain inter-site processes 
(cf.\ section \ref{sec:loc}). Besides stoichiometric compounds with larger primitive cells, 
real-space DFT+DMFT is suited to investigate realistic defect physics in correlated materials, 
e.g., naturally occuring via doping or through intrinsic (point) defects. The research field of 
realistic defect-based many-body physics, albeit a pressing problem in many persistent 
quantum-materials challenges, is only about to receive more attention in recent 
years~\cite{lec14,gri14}.
Let us note that of course also the standard cluster extensions to DMFT (see 
e.g.~\cite{MJPH05} for a review) can be allied with DFT in similar manners. But in the following
subsections we will focus on multi-site applications.

The next two chapters deal with concrete applications using DFT+DMFT calculations based
on the charge self-consistent combination of the mixed-basis pseudopotential 
method~\cite{lou79} for the DFT part and the CT-QMC method, as implemented in 
the TRIQS package~\cite{par15,set16}, for the DMFT impurity problem.

\subsection{The longstanding V$_2$O$_3$ problem as a concrete example
\label{sec:v2o3}}

\begin{figure}[b]
(a)\includegraphics*[height=4.1cm]{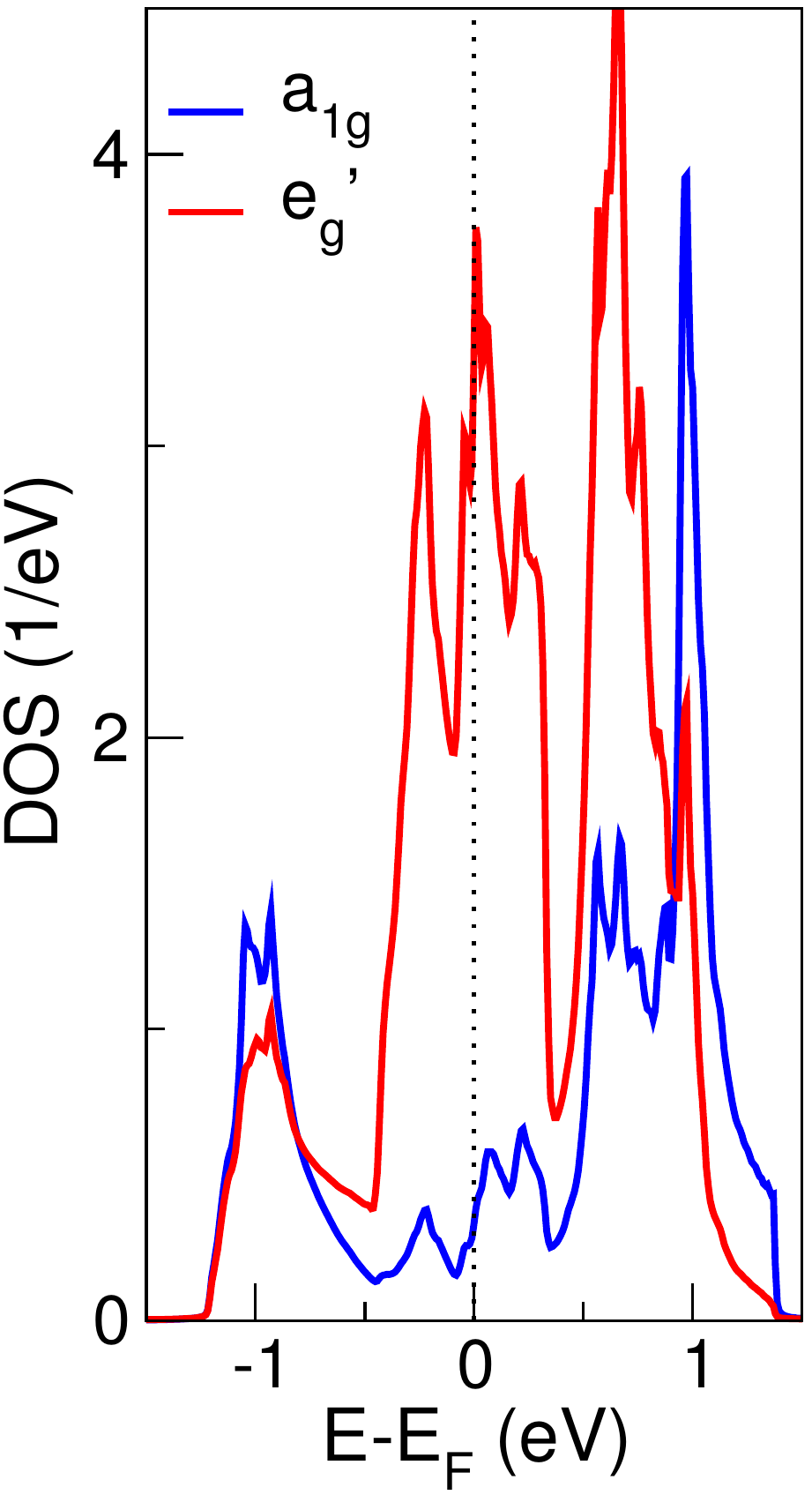}\hspace*{0.2cm}
(b)\includegraphics*[height=4.1cm]{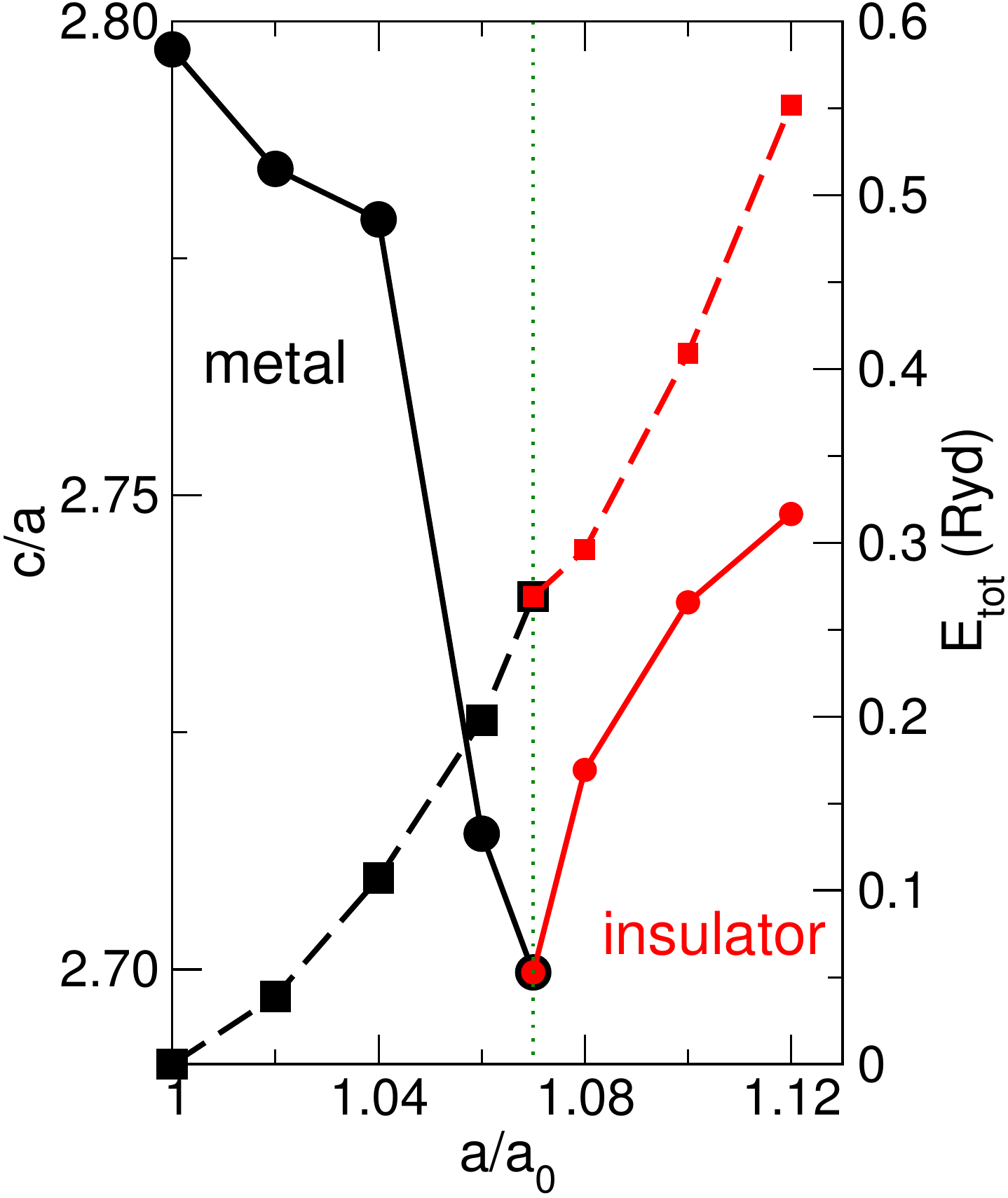}\hspace*{0.2cm}
(c)\includegraphics*[height=4.1cm]{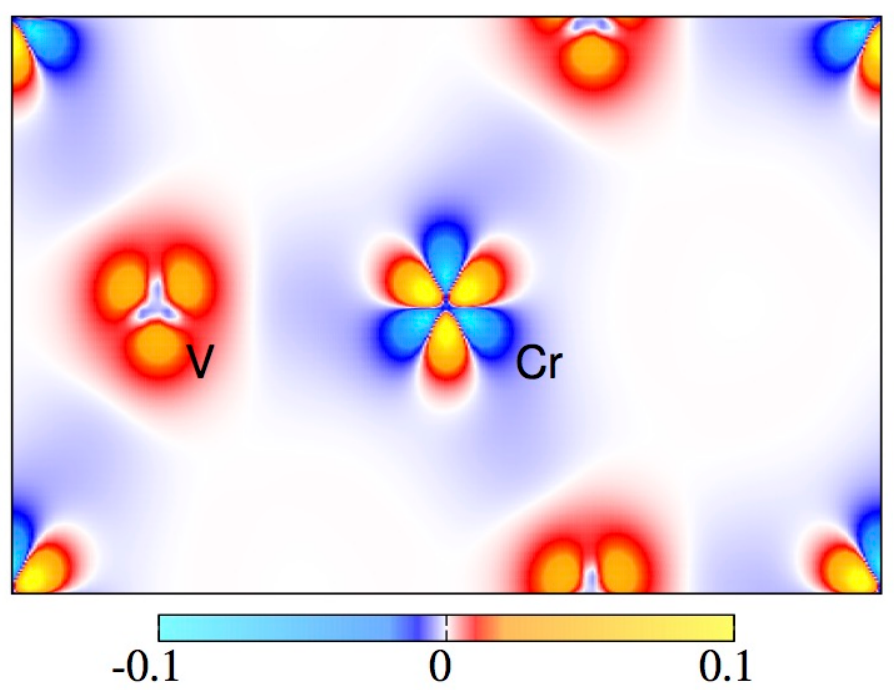}
\caption{Adapted from Refs.\ \cite{gri12,gri14}:
Data for V$_2$O$_3$. 
(a) LDA low-energy density of states. (b) 
Left panel, full lines: relaxed $c/a$ ratio vs. lattice constant.  
Right panel, dashed lines: total energy vs. lattice constant. Black/Red lines mark metal/insulator.
(c) Difference of DFT+DMFT and LDA charge density in the $ab$-plane. Note the
charging of V by Cr due to electronic correlations.}
\label{fig:v2o3}
\end{figure}

For already decades, the vanadium sesquioxide V$_2$O$_3$ poses a famous challenge in the 
understanding of correlated materials~\cite{cas78}. The corundum compound has a canonical phase 
diagram~\cite{mcw69,mcw71} that displays a paramagnetic-metallic, a paramagnetic-insulating as 
well as an antiferromagnetic-insulating  phase at finite $T$, depending 
on pressure or effective electron/hole doping with transition-metal substitutes, i.e., Cr or Ti, 
on the vanadium sites. At ambient $T$ and pressure, the PM phase is the stable one for the 
stoichiometric compound. Along the crystallographic $c$-axis, pairs of V align, and the $ab$-plane
is defined by a honeycomb lattice. A trigonal crystal field acts on the transition-metal site
in VO$_6$ octahedra. Thus the low-energy $t_{2g}$ orbitals of the V($3d$) shell are split into 
an $a_{1g}$ and two degenerate $e_g'$ orbitals, providing a bandwidth $W\sim 2.6$\,eV (see 
Fig.~\ref{fig:v2o3}a). Formally, vanadium is in the oxidation state $3+$,
i.e., a valence configuration $3d^2$ holds. The transition-metal $t_{2g}$ electrons are then also
usually chosen to form the correlated subspace of V$_2$O$_3$. An interacting Hamiltonian of
Slater-Kanamori form, parametrized by a Hubbard $U=5$\,eV and a Hund's exchange 
$J_{\rm H}=0.93$\,eV~\cite{hel01,AKA+12}, is therein applied. Note that the corundum structure asks for
two formula units in its primitive cell, i.e., there are four symmetry-equivalent V ions for
multi-site DFT+DMFT.

From the $U/W$ ratio, the V$_2$O$_3$ system is in a strongly correlated regime, but the
detailed mechanism of the metal-to-insulator transition (MIT) remains a matter of debate. 
Of course, as basic DMFT teaches us, at a large local interaction strength a condensed-matter 
electronic system will become Mott insulating. But this statement is often not sufficient to 
explain or predict details of MITs in materials. The interplay of lattice structure, multi-orbital 
characteristics, defect properties and temperature in a realistic system, allows nature to 
realize quite intricate driving mechanisms. 

%-----------------------------------------------------------------------------------------------------------------------------------------
%-----------------------------------------------------------------------------------------------------------------------------------------

Let us in the following restrict the discussion to the paramagnetic regime. Experimentally, in 
essence, Cr-doping drives V$_2$O$_3$ insulating at constant temperature, accompanied by a subtle 
lattice expansion~\cite{der70} as well as orbital polarization towards $e_g'$~\cite{par00}. 
First-principles DFT~\cite{mat94,ezh99,elf03,eye05,guo14} and 
DFT+DMFT~\cite{hel01,kel04,pot07,gri12,san13,gri14,den14,leo15} studies approached 
the problem. Originally, the focus of most of the theoretical works is on the ``proper averaged''
electronic properties, i.e., leaving the very details of the (electron-)lattice degrees of freedom 
and the doping/defect physics aside. Then, whereas conventional DFT surely fails to describe a MIT, 
``one-shot'' multi-orbital DFT+DMFT indeed suggests a transition. But it does so on the basis of 
experimental lattice data for the metallic and insulating phase~\cite{hel01,kel04,pot07}, while 
keeping the V$_2$O$_3$ stoichiometry in the calculation. The orbital polarization is explained by a 
crystal-field enhancement due to electronic correlations~\cite{kel04,pot07}.

Recently original~\cite{gri12} and follow-up charge self-consistent 
DFT+DMFT~\cite{gri14,den14,leo15} suggests some modifications to this picturing. The 
correlation-enhanced orbital polarization in stoichiometric V$_2$O$_3$ (with or 
without expanded lattice) is predicted to be much weaker. This theoretical forecast
is indeed verified in very recent angle-resolved photoemission experiments~\cite{vec16}. 
Additionally, structural relaxation of the $c/a$ ratio in the correlated system~\cite{gri12} 
exhibits the subtle electron-lattice coupling, marking strong $c/a$ changes close to the MIT 
(see Fig.~\ref{fig:v2o3}b).
Related formal divergences of compressibilities near a Mott transition where before only 
studied within a model context~\cite{has05}. 
A more thorough treating of Cr doping within an explicit supercell approach rendered it furthermore 
obvious that the effect of the chromium dopants may not only be reduced to a sole lattice-expansion 
effect. The valence of Cr has an additional electron and apparently electron-dopes the neighboring 
V sites in an orbital-selective manner (cf. Fig.~\ref{fig:v2o3}c), giving rise to the measured 
orbital polarization~\cite{gri14}.
Note that is has indeed been shown~\cite{rod11}, that applied pressure on insulating Cr-doped 
V$_2$O$_3$ renders the system metallic {\sl without} substantial changes in the orbital polarization.

A complete understanding of the MIT in vanadium sesquioxide is still missing, but DFT+DMFT was so far
very successful in sheding light on various important features. There is
evidence that a full solution asks for the inclusion of correlations, coupling to the lattice and 
defect physics on an close-to equal footing.

\subsection{Oxide Heterostructures: $\delta$-doping of titanate Mott insulators
\label{sec:het}}

The research field of oxide heterostructures has become a prominent one in condensed matter physics 
(see e.g.~\cite{zub11,hwa12,cha14} for reviews). Materials design starting from metallic, 
band- or Mott-insulating transition-metal oxides can lead to emergent physics due to interfacing, 
e.g., to the formation of two-dimensional electron systems. As electronic correlations come easy 
with such oxides, there is the chance of a new era of realistic many-body physics with possibly 
novel exotic phases and various engineering options.

As a highlighting and instructive example, we will show how DFT+DMFT can help to
understand or even predict the complex electronic structure in oxide Mott insulators doped by 
heterostructuring. Experimental work on oxide heterostructures grown
with molecular-beam epitaxy revealed puzzling electron states in rare-earth ($R$) titanates $R$TiO$_3$
upon $\delta$-doping~\cite{jac14,che13}, i.e., monolayer doping, with SrO. The bulk
distorted perovskites are formal Ti$(3d^1)$ systems, whereby the single electron 
occupies the effective $t_{2g}$ subshell. The well-defined $\delta$-doping of such Mott insulators is 
free from disorder effects or screening-length ambiguities. Unusual metallicity as well as complex 
magnetic response has been detected.

%-----------------------------------------------------------------------------------------------------------------------------------------
\begin{figure}[t]
(a)\includegraphics*[height=2.2cm]{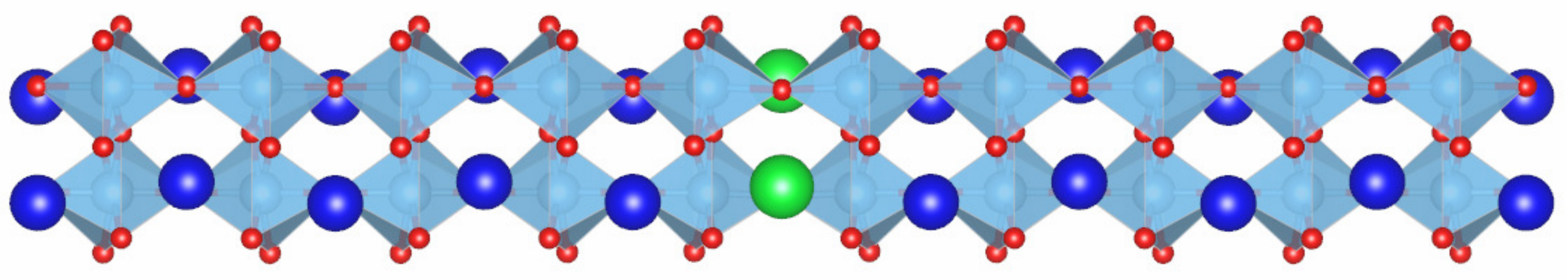}\\[0.1cm]
\parbox[t]{6.75cm}{(b)\hspace*{-0.2cm}\includegraphics*[width=6.4cm]{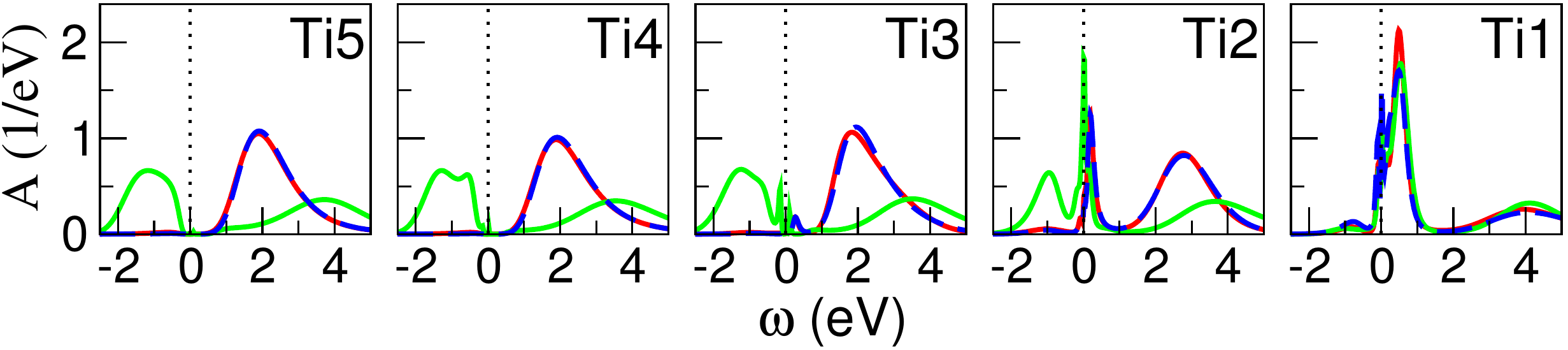}}
\parbox[c]{5.5cm}{(c)\hspace*{-0.2cm}\includegraphics*[width=5.5cm]{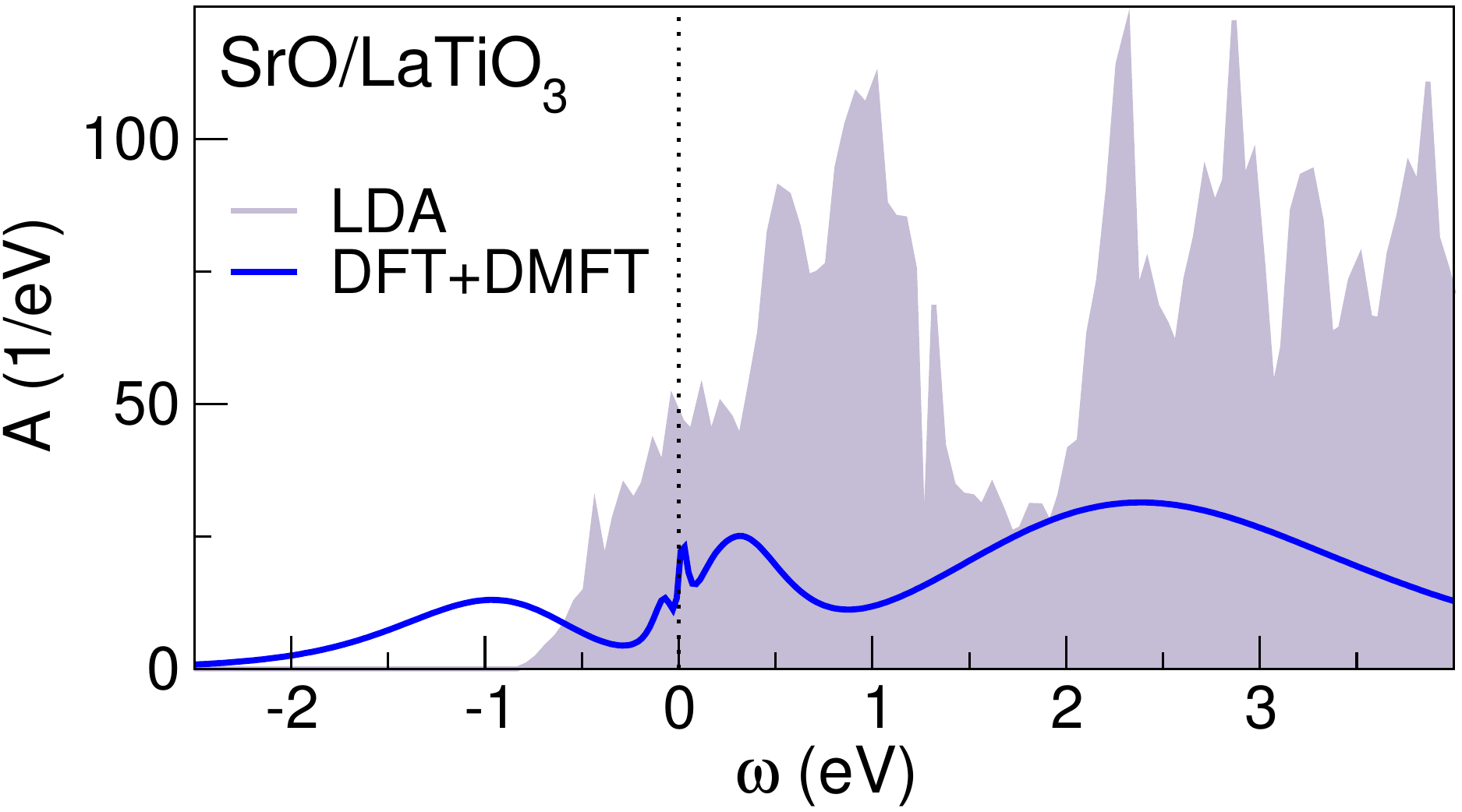}}
\caption{
Adapted from Ref.\ \cite{lec15}:
DFT+DMFT description of $\delta$-doped LaTiO$_3$ ($T=145$\,K). 
(a) Structurally relaxed supercell with La (blue), Ti (grey), O (red) and Sr (green) atoms, 
$c$-axis is aligned horizontally. (b) Local Ti1-5 spectral functions for the effective
$t_{2g}$ orbitals. Ti locations correspond with (a), i.e., Ti1/5 is closest/farest to
the interface. (c) Total spectral function compared to the LDA result.}
\label{fig:delta}
\end{figure}
%-----------------------------------------------------------------------------------------------------------------------------------------

By employing a large supercell (see Fig.~\ref{fig:delta}a) to incorporate the regions close and far away 
from the doping layer, real-space DFT+DMFT calculations including charge self-consistency indeed reveal a 
demanding electronic structure~\cite{lec15}. We focus in the following on the doped LaTiO$_3$ system 
described by a 100-atom primitive cell, with lattice parameters from experiment~\cite{kom07} and
DFT-relaxed atomic positions. There are 20 Ti ions, forming the 5 symmetry-inequivalent classes 
Ti1-5 resembling the different TiO$_2$ layers counted from the SrO doping layer. The correlated subspace 
consists of the effective $t_{2g}$ manifold formed by all the Ti sites and obtained by the 
projected-local-orbital formalism. Again a Slater-Kanamori local-Hamiltonian form is put into practice, 
with $U=5$\,eV and $J_{\rm H}=0.64$\,eV, as proper for titanates~\cite{pav04}.

Of course in DFT, $\delta$-doped LaTiO$_3$ is metallic (as the stoichiometric compound). As displayed in 
Fig.~\ref{fig:delta}c, a metallic state is also detected within DFT+DMFT for the overall 
system, but showing a renormalized quasiparticle peak as well as spectral-weight transfer to a lower
Hubbard band.  Moreover the calculations uncover a rich correlated electronic structure with respect to 
the distance from the SrO doping layer (see Fig.~\ref{fig:delta}b). Far from there, indeed
a Mott-insulating state is stabilized, with nearly complete orbital polarization to one effective
$t_{2g}$ level. The TiO$_2$ layer next to SrO harbors a nearly-orbital-polarized metallic state that
appears Fermi-liquid like~\cite{lec15}. The electrons in the second TiO$_2$ layer between those two 
states form an orbital-polarized doped-Mott state with metallic response but seemingly unusual-large 
electron-electron scattering. In a recent advancement of these investigations to the case of 
$\delta$-doped SmTiO$_3$, a rare-earth titanate close to an antiferromagnetic-ferromagnetic instability,
non-Fermi-liquid behavior and fingerprints of a pseudogap have been revealed~\cite{lec16-2}.

The findings render it obvious that not only a well-defined doped-Mott system with layer-dependent
electron characteristics may be realized by oxide heterostructuring. A further (theoretical) engineering 
of those characteristics appears possible via, e.g., additional chemical dopants, strain, different 
layerings, etc., opening the possibility for the intriguing stabilization of known and/or exotic orders .

%%%%%%%%%%%%%%%%%%%%%%%%%%%%%%%%%%%%%%%%%%%%%%%%%%%
\section{Realistic nonlocal correlations: spin-polaron physics in Na$_x$CoO$_2$}
\label{sec:co}

%-----------------------------------------------------------------------------------------------------------------------------------------
\begin{figure}[t]
\centering
(a)\includegraphics*[height=4.15cm]{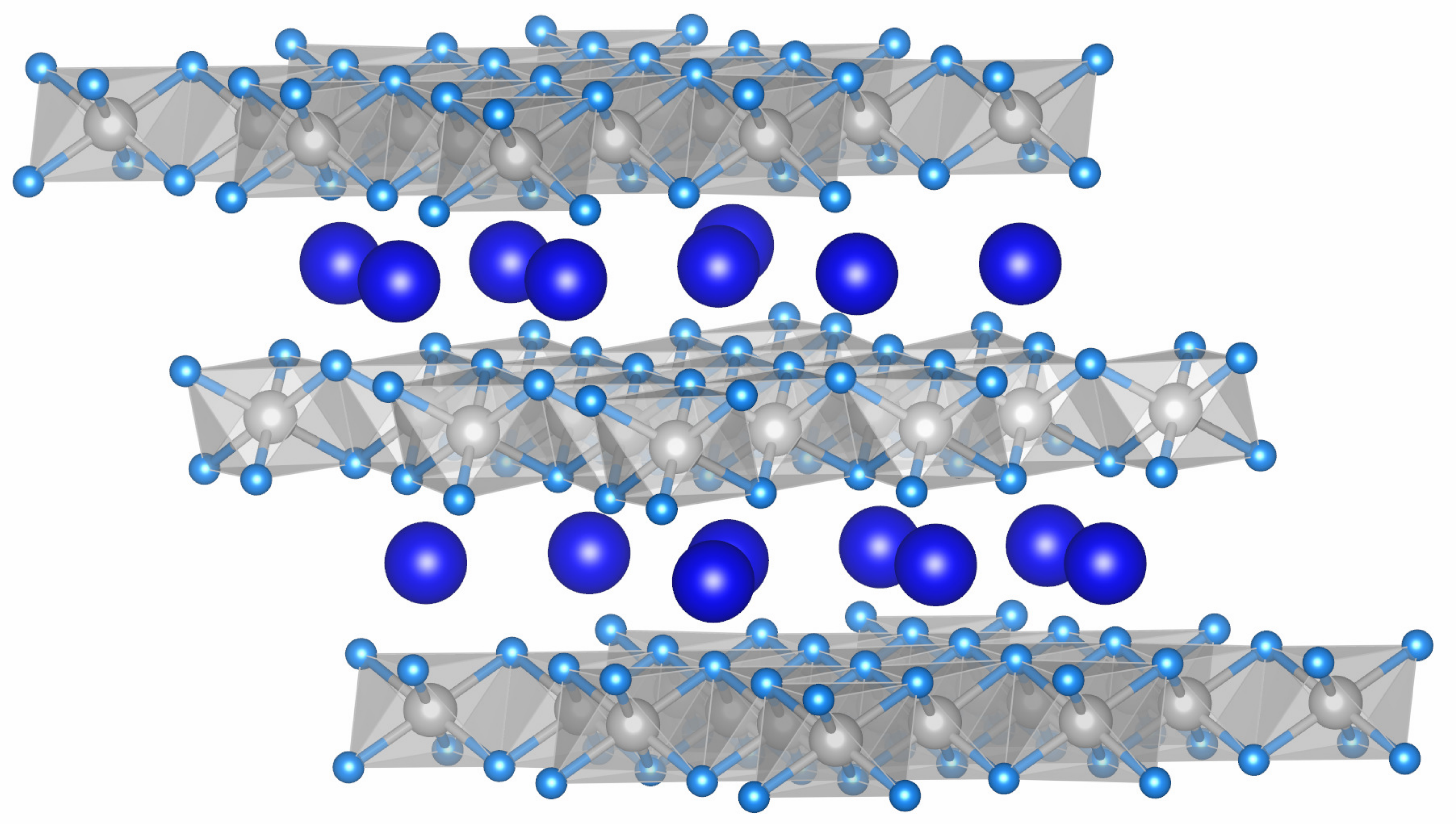}
(b)\includegraphics*[height=4.25cm]{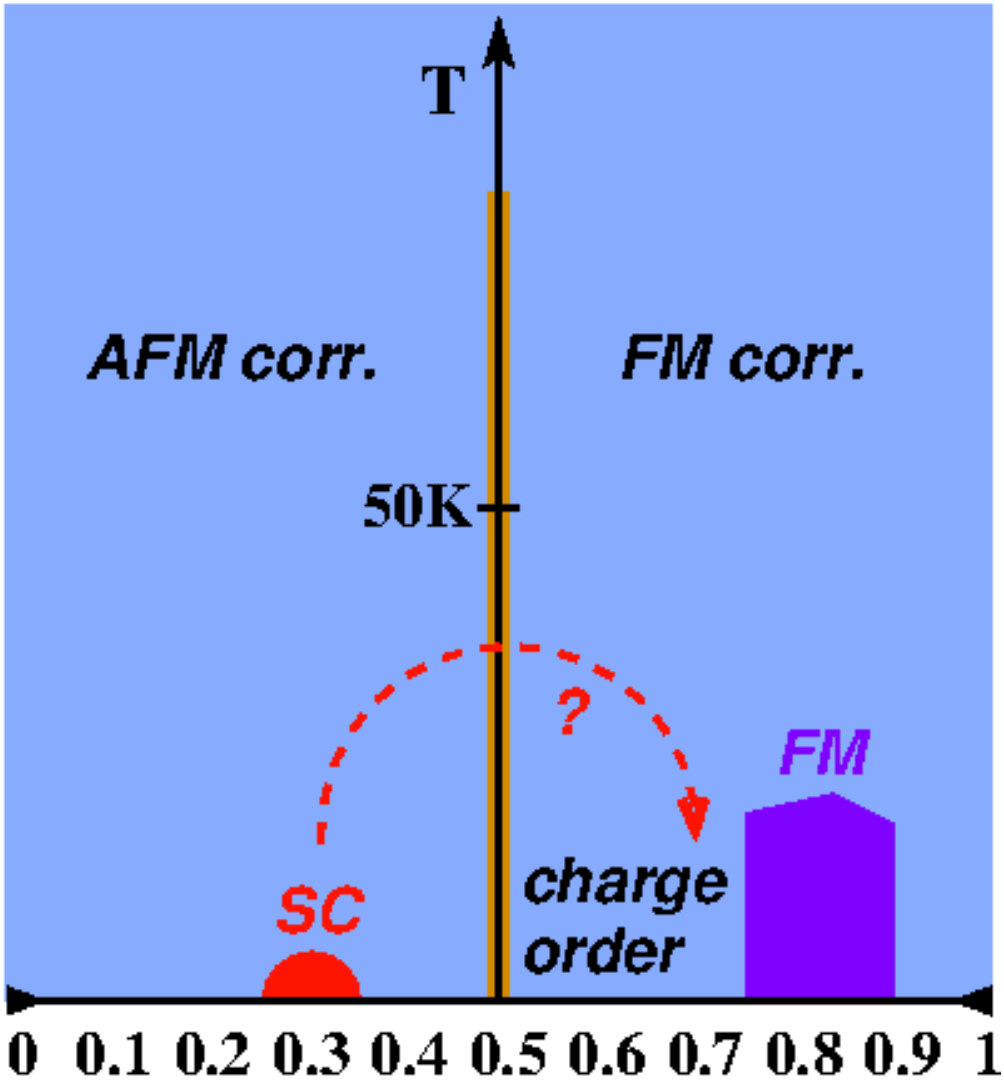}\\[0.4cm]
(c)\includegraphics*[height=4.15cm]{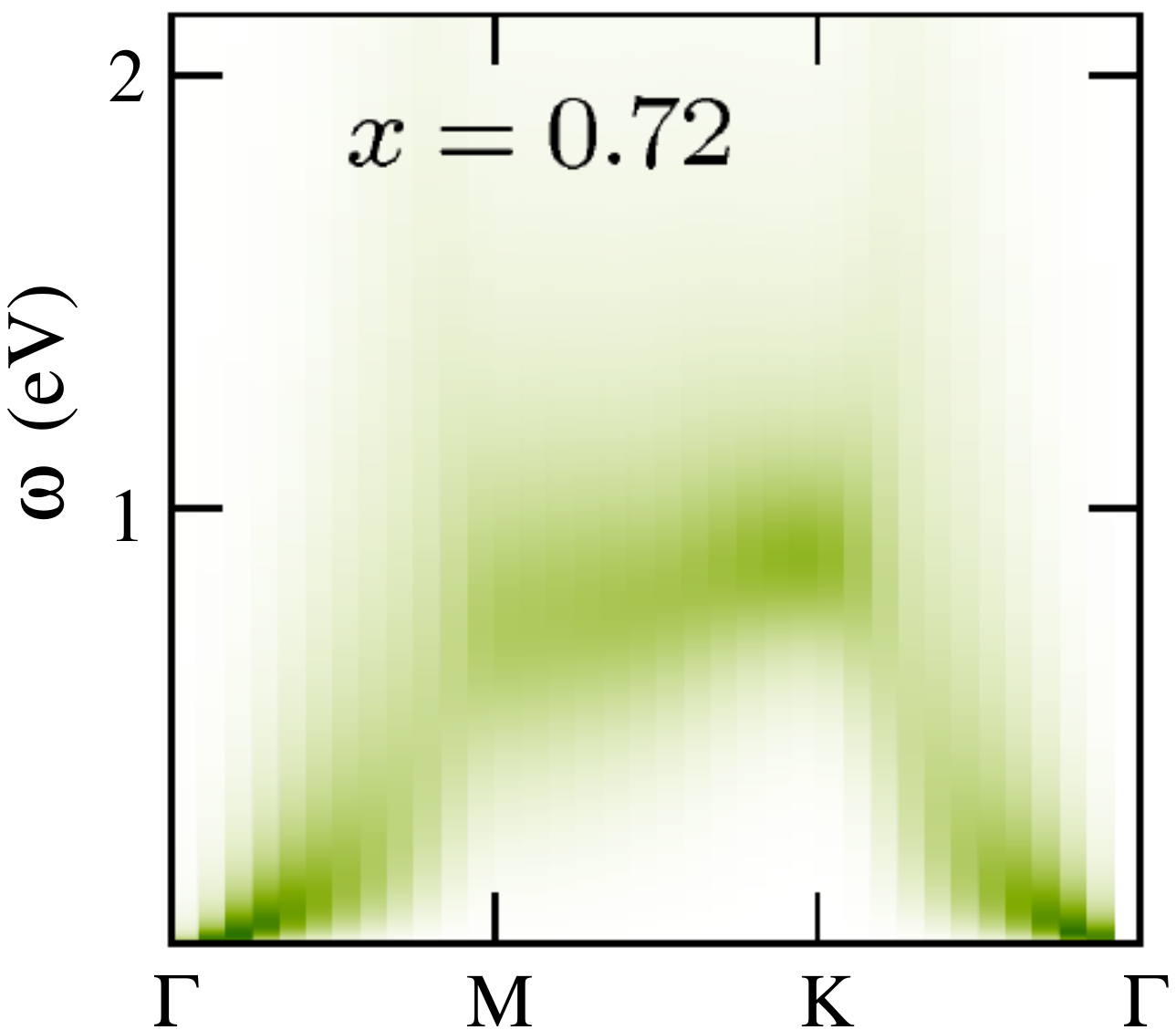}
(d)\includegraphics*[height=4.15cm]{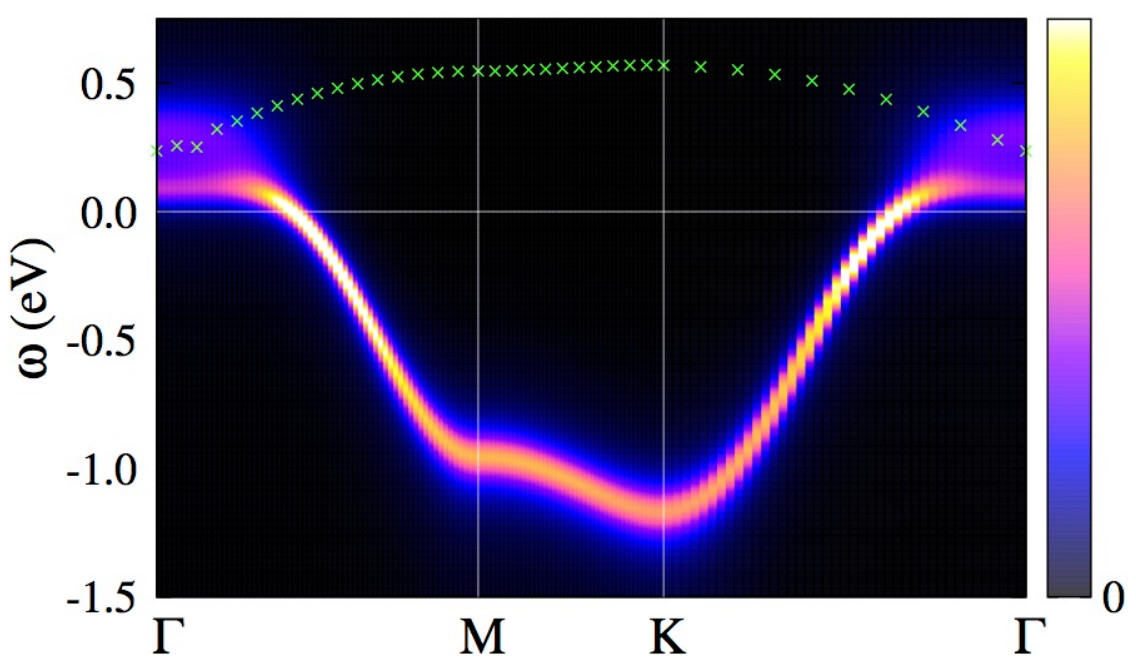}
\caption{
Adapted from Refs.\ \cite{boe12,wil15}:
The Na$_x$CoO$_2$ system harboring electron-paramagnon interactions. 
(a) Crystal structure with Na (large blue), Co (grey) and O (small blue) ions. (b) Sketched $x-T$ in-plane
phase diagram, including the open issue~\cite{oht11} on the location of the superconducting region 
upon water intercalation. (c) DMFT-based dynamic in-plane spin susceptibility obtained by a CT-QMC impurity 
solver for $x=0.72$ including vertex corrections. 
(d) Effective one-band dispersion for $x=0.7$ resulting from the dual-fermion method in the ladder approximation. 
Green crosses mark a model spin-polaron dispersion (see Ref.\ \cite{wil15} for details).}
\label{fig:coba}
\end{figure}
%-----------------------------------------------------------------------------------------------------------------------------------------

So far we discussed nonlocal self-energies $\Sigma({\bf k},\omega)$ in a model context and local 
self-energies $\Sigma(\omega)$ for realistic cases. Due to complexity, realistic nonlocal
correlations are (even) much more demanding. Nonetheless, we are able to report a realistic 
dual-fermion study on the interaction of electrons and paramagnons in strongly correlated sodium 
cobaltate Na$_x$CoO$_2$, close the in-plane ferromagnetic order at $x=0.7$~\cite{wil15}.
Let us briefly summarize the problem and the obtained results. Sodium cobaltate consists of
triangular CoO$_2$ planes subject to strong electronic correlations, that are held together by 
Na ions in between (cf. Fig.~\ref{fig:coba}a). As shown in Fig.~\ref{fig:coba}b, the resulting phase 
diagram upon variation of 
the sodium content $x$ is rather rich, and we are here focussing on the onset of in-plane ferromagnetism 
close to the band-insulating $x=1$ limit. While for smaller $x$ a single hole resides in the 
low-energy Co($t_{2g}$) shell, giving rise to Mott physics, at larger $x$ ferromagnetic fluctuations 
have their share on the electronic correlations. In a first step, we computed $q$-dependent dynamic 
lattice susceptibilities beyond the bubble approximations, i.e., included relevant vertex 
corrections~\cite{boe12}. The latter are based on the two-particle Green's function and the extraction of
the local vertex function~\cite{boe11,boe12}. Figure~\ref{fig:coba}c displays the obtained paramagnon
dispersion close to the $\Gamma$ point in reciprocal space at higher doping $x$. This enhanced
spin susceptibility close to $\Gamma$ should affect the one-particle spectral function, namely
by a coupling between electrons and paramagnons. But as described in section \ref{sec:non}, such 
processes may just be described within the dual-fermion scheme. In short, the $k$-dependent dual 
self-energy can be expressed as
\begin{equation}
\label{sigma}
\tilde{\Sigma}(\bk,\omega) = \sum_{\alpha\nu\omega'\bq} a_{\alpha}
\gamma^{\nu\,\alpha}_{\omega\omega'}\,
\tilde{G}(\bk+\bq,\omega+\nu)\,\tilde{\chi}_{\nu}^{\omega'}(\bq)\,
\hat{\Gamma}^{\nu\,\alpha}_{\omega'\omega}(\bq)\quad,
\end{equation}
whereby $\alpha$ differs spin and charge channel, $\omega$/$\nu$ are fermionic/bosonic Matsubara
frequencies, $\chi$ is the dual particle-hole bubble, $\tilde{G}$ the dual Green's function, $\hat{\Gamma}$
the lattice vertex function and $\gamma$ the local irreducible vertex. Thus it indeed describes the
self-energy building up from one-particle and two-particle objects.

Using a tailored one-band description of the low-energy cobaltate physics, we thus derived a
realistic DFT-based dispersion, applied a proper Hubbard $U=5$\,eV and solved for the local and
nonlocal correlations by the elaborate dual-fermion framework (see~\cite{wil15} for details). And 
indeed, for $x=0.7$ an additional anti-bound state is detected close to $\Gamma$, split off from the 
renormalized quasiparticle dispersion (cf. Fig.~\ref{fig:coba}c). Again in terms of physics, it corresponds
to the interaction of renormalized electrons with strong ferromagnetic fluctuations.
The resulting emerging excitation is a so-called spin polaron and experimental fingerprints thereof
have indeed been detected in optics measurements by Wan {\sl et al.}~\cite{wan04}. Hence besides
refinements of the DMFT-determined electronic spectrum, the inclusion of nonlocal correlations may be
important to reveal more complex excitations with possible relevance for future materials science.

%%%%%%%%%%%%%%%%%%%%%%%%%%%%%%%%%%%%%%%%%%%%%%%%%%%
\section{Perspectives}
\label{sec:con}

The extremely diverse and fascinating physical properties of classical and modern condensed-matter systems, ranging from simple metals (as, e.g., Al) over semiconductors (such as GaAs), itinerant ferromagnets (Fe), Mott or charge-transfer insulators (NiO), to high-$T_{\rm C}$ superconductors (La$_{1-x}$Ba$_{x}$CuO$_{4}$) and graphene, to name a few examples only, are to a large extent determined by their electronic structure. 
It is clear that a realistic modeling of the interacting electron gas in a solid that covers materials of such different kind in a coherent theoretical framework represents a highly ambitious project. 
In fact, this project goes on for decades and is by no means finished. 
It is triggered by an ever-improving set of sophisticated experimental techniques, in particular by the various breakthroughs in high-resolution photoemission spectroscopy, which is the technique that gives the most direct access to the electronic structure. 
Also the discovery of new materials or the completely revised understanding of ``well-known'' ones pushes this project.
Predictive power is the regulative idea, which drives the research in this field, since the ultimate goal of any theory should be to make measurable predictions, as precise as possible, without any {\em a posteriori} input, and starting only from general information, such as chemical composition.

With the focus on strongly correlated electron systems, the next level in this endeavor is reached. 
Typically, the electronic properties of these correlated materials are extremely susceptible to all kinds of perturbations. 
This may lead to a substantially increased complexity of the essential physics.
A related characteristics of strongly correlated systems is that different electronic mechanisms are at work, favoring phases with different types of local, short- or lang-range correlations. 
These compete or cooperate with each other at low temperatures and on tiny energy scales. 
Besides the bare energy scales, given by the one-particle or interaction parameters, also new and typically much smaller energy scales are generated by correlations. 
This can give rise to involved thermodynamical phase diagrams, already for model systems as we have seen in section \ref{sec:loc} where single-site screening effects, nonlocal magnetic correlations and geometrical frustration compete with each other.

Understanding strongly correlated systems thus requires (i) nonperturbative approaches to tackle the correlation problem and (ii) highly accurate tools to describe the chemical and geometrical aspects and to compute one-particle and interaction matrix elements.
Using dynamical mean-field theory, in a proper combination with refined concepts of band-structure theory, substantial progress could be made in the recent past.

While the separate potentials of DFT and of DMFT themselves are not at all fully exhausted, the versatile usability and the robustness of the band-structure and of the mean-field concept make the combination of DFT with DMFT an almost ideal approach to cover complex materials with strong correlations.
Examples for involved multi-orbital physics have been discussed in sections \ref{sec:v2o3} and \ref{sec:co}. 
In this context, there are several possibly important issues that must be accounted for and that can be covered in fact, such as low-symmetry crystal structures, intricate crystal-field effects, structural distortions, sophisticated screening processes, etc.
This also includes lack of translational symmetry, see section \ref{sec:het}. 
Inhomogeneous correlated systems, such as surfaces and interfaces, can be studied with the real-space variant of DFT+DMFT (section \ref{sec:ddrealspace}).
This is important to make close quantitative contact with highly surface-sensitive photoemission spectroscopy. 

To advance DMFT to a first-principles approach with predictive power for real materials, spanning traditionally studied and modern strongly correlated materials, special attention must be spent on the interface with density-functional theory. 
The interface is crucial as the nonlocal concepts from band theory must be combined with the locality paradigm of DMFT. 
Related to this is the choice of a proper one-particle basis set, the definition of a physically motivated ``correlated subspace'' and the forth and back switching (``up and down folding'') between the DMFT treatment of lattice model with local interactions in the correlated subspace on the one hand, and the effective one-particle multiple-scattering problem in the entire orbital space on the other.
Actually, this full complexity shows up when running through the ``big self-consistency loop''. 
This comes in addition to the internal loops that are characteristic for DMFT, which self-consistently maps onto an effective impurity problem, and for DFT, which self-consistently maps onto the homogeneous electron gas. 
The big loop or charge-self-consistency provides the necessary feedback of the correlation-modified charge density on the effective one-particle potential (cf.\ section \ref{sec:ddbasics}).
Also in this respect, we have witnessed great progress in the recent years.

What is the to-do list for future developments?
This question has at least two lines of possible answers. 
First, the present activities must be continued to further evolve the DFT+DMFT approach. 
This includes the development of better impurity solvers, which are able to access lower temperatures and to treat impurity models with more and more orbitals, and which give reliable real-frequency information.
Improved quantum Monte-Carlo techniques (along the lines of Ref.\ \cite{GML+11}), approaches based on the time-dependent density-matrix renormalization group \cite{Sch11} or independent novel ideas \cite{LHGH14} are promising. 
Second, extensions of the theory which incorporate the feedback of nonlocal correlations on the self-energy must be advanced (see our discussion in section \ref{sec:nonloc}). 
In addition, further extensions are necessary to treat systems with nonlocal interactions. 
There are different diagrammatic or Green's-function-based approaches which are worth pursuing (cf.\ section \ref{sec:nonloc}), but it is probably fair to say that the present theoretical ideas are far from providing an answer that is  conceptually as clear as DMFT has given to the question of how to treat local correlations and local interactions. 
A third problem is the interface to the experiment. 
Besides static thermodynamical properties, the central quantity of DFT+DMFT is the one-particle spectral function which, however, is not directly observable. 
A quantitative prediction of photoemission spectra rather requires a proper inclusion of transition-matrix elements, an important issue that is in most cases set aside, unnecessarily. 
A challenge, on the other hand, is the development of a realistic theory of time-dependent (pump-probe) photoemission from correlated materials (see Ref.\ \cite{BRP+15} for first attempts). 
While nonequilibrium DMFT \cite{ATE+14} is already on the market, a nonequilibrium DFT+DMFT, however, is not yet in sight.

Finally, there are some future challenges which are of a more fundamental character: 
One example is the so-called double-counting problem, i.e., an incorrect double counting of interactions, once on some average level within the local-density approximation of DFT or variants of the LDA and once more, and more explicit, in the correlated lattice model to which DMFT is applied. 
While double-counting ``corrections'' are regularly used in practical DFT+DMFT calculations and while in most cases this is fully sufficient from a pragmatic point of view, all of them lack a thorough theoretical foundation.
The GW+DMFT approach elegantly circumvents the double-counting problem, but eliminates the DFT-part of DFT+DMFT altogether, and it is surely questionable whether GW as a starting point is as reliable and versatile as the more traditional DFT. 

Another fundamental critique targets the DMFT as such, and also its cluster or diagrammatic extensions, as these are mean-field approaches in the end and thus also share some typical problems associated with the mean-field concept.
Critical phenomena, for example, are not easily described correctly when starting from the local DMFT. 
Also the possibility for nonunique solutions, resulting from the nonlinearity of the DMFT self-consistency (or mean-field) equation, must be seen as problematic from a fundamental quantum-statistical point of view - although this is often seen as advantageous and exploited in practice. 
Finally, it is by no means obvious that the central quantity of DMFT, the one-particle Green's function, deserves a distinguished role. 
Competing theories start from different observables, such as higher-order Green's functions, or even the many-body wave function, constrained, e.g., to some ``physical corner'' of the Hilbert space \cite{GE16}, or the reduced one-particle density matrix. 
First attempts to clarify the mutual relation between the different concepts appear very promising \cite{BPP13}.
Up to now, however, there is no working alternative to the DFT+DMFT concept when addressing the fascinating world of correlation effects in real materials.

\acknowledgement
We would like to thank M.\ W.\ Aulbach, L.\ Boehnke, D.\ Grieger, H.\ Hafermann and E.\ G.\ C.\ P.\ van Loon for the intense and fruitful cooperation over the years. 
Financial support of this work by the Deutsche Forschungsgemeinschaft through the Forschergruppe FOR 1346 (project P1) is gratefully acknowledged.


\begin{thebibliography}{113}

\bibitem{MJW71}
P.M. Marcus, J.F. Janak, A.R. Williams, eds., \emph{Computational Methods in
  Band Theory} (Plenum, New York, 1971)

\bibitem{And75}
O.K. Andersen, Phys. Rev. B \textbf{12}, 3060 (1975)

\bibitem{BSST90}
P.~Blaha, K.~Schwarz, P.~Sorantin, S.B. Trickey, Comput. Phys. Commun.
  \textbf{59}, 399 (1990)

\bibitem{BDGW92}
W.H. Butler, P.H. Dederichs, A.~Gonis, R.L. Weaver, \emph{Applications of
  Multiple Scattering Theory to Material Science} (Materials Reserach Society,
  Pittsburg Penn., 1992)

\bibitem{HK64}
P.~Hohenberg, W.~Kohn, Phys. Rev. \textbf{136}, 864 (1964)

\bibitem{KS65}
W.~Kohn, L.J. Sham, Phys. Rev. \textbf{140}, 1133 (1965)

\bibitem{JG89}
R.O. Jones, O.~Gunnarsson, Rev. Mod. Phys. \textbf{61}, 689 (1989)

\bibitem{DDN98}
M.~Donath, P.A. Dowben, W.~Nolting, eds., \emph{Magnetism and electronic
  correlations in local-moment systems: Rare-earth elements and compounds}
  (World Scientific, Singapore, 1998)

\bibitem{Mot90}
N.F. Mott, \emph{Metal-Insulator Transitions}, 2nd~edn. (Taylor and Francis,
  London, 1990)

\bibitem{OM00}
J.~Orenstein, A.J. Millis, Science \textbf{288}, 468 (2000)

\bibitem{AGD64}
A.A. Abrikosow, L.P. Gorkov, I.E. Dzyaloshinski, \emph{Methods of Quantum Field
  Theory in Statistical Physics} (Prentice-Hall, New Jersey, 1964)

\bibitem{Huf07}
S.~H\"ufner, ed., \emph{Very high resolution photoelectron spectroscopy}
  (Springer, New York, 2007)

\bibitem{Gut63}
M.C. Gutzwiller, Phys. Rev. Lett. \textbf{10}, 159 (1963)

\bibitem{Hub63}
J.~Hubbard, Proc. R. Soc. London A \textbf{276}, 238 (1963)

\bibitem{Kan63}
J.~Kanamori, Prog. Theor. Phys. (Kyoto) \textbf{30}, 275 (1963)

\bibitem{MV89}
W.~Metzner, D.~Vollhardt, Phys. Rev. Lett. \textbf{62}, 324 (1989)

\bibitem{GK92a}
A.~Georges, G.~Kotliar, Phys. Rev. B \textbf{45}, 6479 (1992)

\bibitem{GKKR96}
A.~Georges, G.~Kotliar, W.~Krauth, M.J. Rozenberg, Rev. Mod. Phys. \textbf{68},
  13 (1996)

\bibitem{KV04}
G.~Kotliar, D.~Vollhardt, Physics Today \textbf{57}, 53 (2004)

\bibitem{CK94}
M.~Caffarel, W.~Krauth, Phys. Rev. Lett. \textbf{72}, 1545 (1994)

\bibitem{BCP08}
R.~Bulla, T.A. Costi, T.~Pruschke, Rev. Mod. Phys. \textbf{80}, 395 (2008)

\bibitem{GML+11}
E.~Gull, A.~Millis, A.~Lichtenstein, A.~Rubtsov, M.~Troyer, P.~Werner, Rev.
  Mod. Phys. \textbf{83}, 349 (2011)

\bibitem{GTV+14}
M.~Ganahl, P.~Thunstr\"om, F.~Verstraete, K.~Held, H.G. Evertz, Phys. Rev. B
  \textbf{90}, 045144 (2014)

\bibitem{LHGH14}
Y.~Lu, M.~H\"oppner, O.~Gunnarsson, M.W. Haverkort, Phys. Rev. B \textbf{90},
  085102 (2014)

\bibitem{MJPH05}
T.~Maier, M.~Jarrell, T.~Pruschke, M.H. Hettler, Rev. Mod. Phys. \textbf{77},
  1027 (2005)

\bibitem{TKH07}
A.~Toschi, A.A. Katanin, K.~Held, Phys. Rev. B \textbf{75}, 045118 (2007)

\bibitem{RKL08}
A.N. Rubtsov, M.I. Katsnelson, A.I. Lichtenstein, Phys. Rev. B \textbf{77},
  033101 (2008)

\bibitem{RKL12}
A.N. Rubtsov, M.I. Katsnelson, A.I. Lichtenstein, Ann. Phys. \textbf{327}, 1320
  (2012)

\bibitem{Anderson61}
P.W. Anderson, Phys. Rev. \textbf{124}, 41 (1961)

\bibitem{Fisk_rev}
Z.~Fisk, H.~Ott, T.M.Rice, J.~Smith, Nature \textbf{320}, 124 (1986)

\bibitem{Tsunetsugu97_rev}
H.~Tsunetsugu, M.~Sigrist, K.~Ueda, Rev. Mod. Phys. \textbf{69}, 809 (1997)

\bibitem{Capponi00}
S.~Capponi, F.F. Assaad, Phys. Rev. B \textbf{63}, 155114 (2001)

\bibitem{SW66}
J.R. Schrieffer, P.A. Wolff, Phys. Rev. \textbf{149}, 491 (1966)

\bibitem{Don77}
S.~Doniach, Physica \textbf{91B}, 321 (1977)

\bibitem{Lohneysen_rev}
H.~v.~L\"ohneysen, A.~Rosch, M.~Vojta, P.~W\"olfle, Rev. of Mod. Phys.
  \textbf{79}, 1015 (~61) (2007)

\bibitem{RK54}
M.A. Ruderman, C.~Kittel, Phys. Rev. \textbf{96}, 99 (1954)

\bibitem{Kondo64}
J.~Kondo, Prog. Theor. Phys. \textbf{32}, 37 (1964)

\bibitem{Hewson}
A.C. Hewson, \emph{The Kondo Problem to Heavy Fermions}, Cambridge Studies in
  Magnetism (Cambridge Universiy Press, Cambridge, 1997)

\bibitem{AAP15}
M.W. Aulbach, F.F. Assaad, M.~Potthoff, Phys. Rev. B \textbf{92}, 235131 (2015)

\bibitem{RSL05}
A.N. Rubtsov, V.V. Savkin, A.I. Lichtenstein, Phys. Rev. B \textbf{72}, 035122
  (2005)

\bibitem{WCM+06}
P.~Werner, A.~Comanac, L.~de' Medici, M.~Troyer, A.J. Millis, Phys. Rev. Lett.
  \textbf{97}, 076405 (2006)

\bibitem{MNY+10}
Y.~Motome, K.~Nakamikawa, Y.~Yamaji, M.~Udagawa, Phys. Rev. Lett. \textbf{105},
  036403 (2010)

\bibitem{Ballou91}
R.~Ballou, C.~Lacroix, M.D. Nunez~Regueiro, Phys. Rev. Lett. \textbf{66}, 1910
  (1991)

\bibitem{BFV11}
A.~Benlagra, L.~Fritz, M.~Vojta, Phys. Rev. B \textbf{84}, 075126 (2011)

\bibitem{HUM11}
S.~Hayami, M.~Udagawa, Y.~Motome, J. Phys. Soc. Jpn. \textbf{80}, 073704 (2011)

\bibitem{HUM12}
S.~{Hayami}, M.~{Udagawa}, Y.~{Motome}, J. Phys. Soc. Jpn. \textbf{81}, 103707
  (2012)

\bibitem{PN97b}
M.~Potthoff, W.~Nolting, Phys. Rev. B \textbf{55}, 2741 (1997)

\bibitem{TSRP12}
I.~Titvinidze, A.~Schwabe, N.~Rother, M.~Potthoff, Phys. Rev. B \textbf{86},
  075141 (2012)

\bibitem{ATP15}
M.W. Aulbach, I.~Titvinidze, M.~Potthoff, Phys. Rev. B \textbf{91}, 174420
  (2015)

\bibitem{kotliar_review}
G.~Kotliar, S.Y. Savrasov, K.~Haule, V.S. Oudovenko, O.~Parcollet, C.A.
  Marianetti, Rev. Mod. Phys. \textbf{78}, 865 (2006)

\bibitem{Biermann03}
S.~Biermann, F.~Aryasetiawan, A.~Georges, Phys. Rev. Lett. \textbf{90}, 086402
  (2003)

\bibitem{Sun02}
P.~Sun, G.~Kotliar, Phys. Rev. B \textbf{66}, 085120 (2002)

\bibitem{vanLoon2014}
E.G.C.P. van Loon, H.~Hafermann, A.I. Lichtenstein, A.N. Rubtsov, M.I.
  Katsnelson, Phys. Rev. Lett. \textbf{113}, 246407 (2014)

\bibitem{Stepanov2016}
E.A. Stepanov, E.G.C.P. van Loon, A.A. Katanin, A.I. Lichtenstein, M.I.
  Katsnelson, A.N. Rubtsov, Phys. Rev. B \textbf{93}, 045107 (2016)

\bibitem{ani97}
V.I. Anisimov, A.I. Poteryaev, M.A. Korotin, A.O. Anokhin, G.~Kotliar, J.
  Phys.: Condens. Matter \textbf{9}, 7359 (1997)

\bibitem{lic98}
A.I. Lichtenstein, M.~Katsnelson, Phys. Rev. B \textbf{57}, 6884 (1998)

\bibitem{min05}
J.~Min\'{a}r, L.~Chioncel, A.~Perlov, H.~Ebert, M.I. Katsnelson, A.I.
  Lichtenstein, Phys. Rev. B \textbf{72}, 045125 (2005)

\bibitem{mar12}
N.~Marzari, A.A. Mostofi, J.R. Yates, I.~Souza, D.~Vanderbilt, Rev. Mod. Phys.
  \textbf{84}, 1419 (2012)

\bibitem{lec06}
F.~Lechermann, A.~Georges, A.~Poteryaev, S.~Biermann, M.~Posternak,
  A.~Yamasaki, O.K. Andersen, Phys. Rev. B \textbf{74}, 125120 (2006)

\bibitem{ama08}
B.~Amadon, F.~Lechermann, A.~Georges, F.~Jollet, T.O. Wehling, A.I.
  Lichtenstein, Phys. Rev. B \textbf{77}, 205112 (2008)

\bibitem{ani05}
V.I. Anisimov, D.E. Kondakov, A.V. Kozhevnikov, I.A. Nekrasov, Z.V. Pchelkina,
  J.W. Allen, S.K. Mo, H.D. Kim, P.~Metcalf, S.~Suga et~al., Phys. Rev. B
  \textbf{71}, 125119 (2005)

\bibitem{hel01}
K.~Held, G.~Keller, V.~Eyert, D.~Vollhardt, V.I. Anisimov, Phys. Rev. Lett.
  \textbf{86}, 5345 (2001)

\bibitem{lic01}
A.I. Lichtenstein, M.~Katsnelson, G.~Kotliar, Phys. Rev. Lett. \textbf{87},
  067205 (2001)

\bibitem{pav04}
E.~Pavarini, S.~Biermann, A.~Poteryaev, A.I. Lichtenstein, A.~Georges, O.K.
  Andersen, Phys. Rev. Lett. \textbf{92}, 176403 (2004)

\bibitem{sav01}
S.Y. Savrasov, G.~Kotliar, E.~Abrahams, Nature \textbf{410}, 793 (2001)

\bibitem{pou07}
L.V. Pourovskii, B.~Amadon, S.~Biermann, A.~Georges, Phys. Rev. B \textbf{76},
  235101 (2007)

\bibitem{gri12}
D.~Grieger, C.~Piefke, O.E. Peil, F.~Lechermann, Phys. Rev. B \textbf{86},
  155121 (2012)

\bibitem{ani-u91}
V.I. Anisimov, J.~Zaanen, O.K. Andersen, Phys. Rev. B \textbf{44}, 943 (1991)

\bibitem{ani93}
V.I. Anisimov, I.V. Solovyev, M.A. Korotin, M.T. Czy$\dot{z}$yk, G.A. Sawatzky,
  Phys. Rev. B \textbf{48}, 16929 (1993)

\bibitem{czy94}
M.T. Czy$\dot{z}$yk, G.A. Sawatzky, Phys. Rev. B \textbf{49}, 14211 (1994)

\bibitem{geo04}
A.~Georges, \emph{Strongly correlated electron materials: Dynamical mean-field
  theory and electronic structure}, Lectures on the Physics of Highly
  Correlated Electron Systems VIII, AIP Conference Proceedings, Vol. 715, p.3
  (2004)

\bibitem{lec14}
F.~Lechermann, L.~Boehnke, D.~Grieger, C.~Piefke, Phys. Rev. B \textbf{90},
  085125 (2014)

\bibitem{gri14}
D.~Grieger, F.~Lechermann, Phys. Rev. B \textbf{90}, 115115 (2014)

\bibitem{lou79}
S.G. Louie, K.M. Ho, M.L. Cohen, Phys. Rev. B \textbf{19}, 1774 (1979)

\bibitem{par15}
O.~Parcollet, M.~Ferrero, T.~Ayral, H.~Hafermann, I.~Krivenko, L.~Messio,
  P.~Seth, Comput. Phys. Commun. \textbf{196}, 398 (2015)

\bibitem{set16}
P.~Seth, I.~Krivenko, M.~Ferrero, O.~Parcollet, Comput. Phys. Commun.
  \textbf{200}, 274 (2016)

\bibitem{cas78}
C.~Castellani, C.R. Natoli, J.~Ranninger, Phys. Rev. B \textbf{18}, 4945 (1978)

\bibitem{mcw69}
D.B. McWhan, T.M. Rice, J.B. Remeika, Phys. Rev. Lett. \textbf{23}, 1384 (1969)

\bibitem{mcw71}
D.B. McWhan, J.B. Remeika, T.M. Rice, W.F. Brinkman, J.P. Maita, A.~Menth,
  Phys. Rev. Lett. \textbf{27}, 941 (1971)

\bibitem{AKA+12}
A.E. Antipov, I.S. Krivenko, V.I. Anisimov, A.I. Lichtenstein, A.N. Rubtsov,
  Phys. Rev. B \textbf{86}, 155107 (2012)

\bibitem{der70}
P.D. Dernier, J. Phys. Chem. Solids \textbf{31}, 2569 (1970)

\bibitem{par00}
J.H. Park, L.H. Tjeng, A.~Tanaka, J.W. Allen, C.T. Chen, P.~Metcalf, J.M.
  Honig, F.M.F. de~Groot, G.A. Sawatzky, Phys. Rev. B \textbf{61}, 11506 (2000)

\bibitem{mat94}
L.F. Mattheiss, J. Phys.: Condens. Matter \textbf{6}, 6477 (1994)

\bibitem{ezh99}
S.Y. Ezhov, V.I. Anisimov, D.I. Khomskii, G.A. Sawatzky, Phys. Rev. Lett.
  \textbf{83}, 4136 (1999)

\bibitem{elf03}
I.S. Elfimov, T.~Saha-Dasgupta, M.A. Korotin, Phys. Rev. B \textbf{68}, 113105
  (2003)

\bibitem{eye05}
V.~Eyert, U.~Schwingenschl\"ogl, U.~Eckern, Europhys. Lett. \textbf{70}, 782
  (2005)

\bibitem{guo14}
Y.~Guo, S.J. Clark, J.~Robertson, J. Chem. Phys. \textbf{140} (2014)

\bibitem{kel04}
G.~Keller, K.~Held, V.~Eyert, D.~Vollhardt, V.I. Anisimov, Phys. Rev. B
  \textbf{70}, 205116 (2004)

\bibitem{pot07}
A.I. Poteryaev, J.M. Tomczak, S.~Biermann, A.~Georges, A.I. Lichtenstein, A.N.
  Rubtsov, T.~Saha-Dasgupta, O.K. Andersen, Phys. Rev. B \textbf{76}, 085127
  (2007)

\bibitem{san13}
M.~Sandri, M.~Capone, M.~Fabrizio, Phys. Rev. B \textbf{87}, 205108 (2013)

\bibitem{den14}
X.~Deng, A.~Sternbach, K.~Haule, D.~Basov, G.~Kotliar, Phys. Rev. Lett.
  \textbf{113}, 246404 (2014)

\bibitem{leo15}
I.~Leonov, V.I. Anisimov, D.~Vollhardt, Phys. Rev. B \textbf{91}, 195115 (2015)

\bibitem{vec16}
I.~{Lo Vecchio}, J.D. Denlinger, O.~Krupin, B.J. Kim, P.A. Metcalf, S.~Lupi,
  J.W. Allen, A.~Lanzara, Phys. Rev. Lett. \textbf{117}, 166401 (2016)

\bibitem{has05}
S.R. Hassan, A.~Georges, H.R. Krishnamurthy, Phys. Rev. Lett. \textbf{94},
  036402 (2005)

\bibitem{rod11}
F.~Rodolakis, J.P. Rueff, M.~Sikora, I.~Alliot, J.P. Iti{\'e}, F.~Baudelet,
  S.~Ravy, P.~Wzietek, P.~Hansmann, A.~Toschi et~al., Phys. Rev. B \textbf{84},
  245113 (2011)

\bibitem{zub11}
P.~Zubko, S.~Gariglio, M.~Gabay, P.~Ghosez, J.M. Triscone, Annu. Rev. Condens.
  Matter Phys. \textbf{2}, 141 (2011)

\bibitem{hwa12}
H.Y. Hwang, Y.~Iwasa, M.~Kawasaki, B.~Keimer, N.~Nagaosa, Y.~Tokura, Nature
  Materials \textbf{11}, 103 (2012)

\bibitem{cha14}
J.~Chakhalian, J.W. Freeland, A.J. Millis, C.~Panagopoulos, J.M. Rondinelli,
  Rev. Mod. Phys. \textbf{86}, 1189 (2014)

\bibitem{jac14}
C.A. Jackson, J.Y. Zhang, C.R. Freeze, S.~Stemmer, Nat. Commun. \textbf{5},
  4258 (2014)

\bibitem{che13}
R.~Chen, S.B. Lee, L.~Balents, Phys. Rev. B \textbf{87}, 161119(R) (2013)

\bibitem{lec15}
F.~Lechermann, M.~Obermeyer, New J. Phys. \textbf{17}, 043026 (2015)

\bibitem{kom07}
A.C. Komarek, H.~Roth, M.~Cwik, W.D. Stein, J.~Baier, M.~Kriener,
  F.~Bour{\'e}e, T.~Lorenz, M.~Braden, Phys. Rev. B \textbf{75}, 224402 (2007)

\bibitem{lec16-2}
F.~Lechermann, arXiv:1603.01031  (2016)

\bibitem{boe12}
L.~Boehnke, F.~Lechermann, Phys. Rev. B \textbf{85}, 115128 (2012)

\bibitem{wil15}
A.~Wilhelm, F.~Lechermann, H.~Hafermann et~al., Phys. Rev. B \textbf{91},
  155114 (2015)

\bibitem{oht11}
H.~Ohta, K.~Yoshimura, Z.~Hu, Y.Y. Chin, H.J. Lin, H.H. Hsieh, C.T. Chen, L.H.
  Tjeng, Phys. Rev. Lett. \textbf{107}, 066404 (2011)

\bibitem{boe11}
L.~Boehnke, H.~Hafermann, M.~Ferrero, F.~Lechermann, O.~Parcollet, Phys. Rev. B
  \textbf{84}, 075145 (2011)

\bibitem{wan04}
N.L. Wang, P.~Zheng, D.~Wu, Y.C. Ma, T.~Xiang, R.Y. Jin, D.~Mandrus, Phys. Rev.
  Lett. \textbf{93}, 237007 (2004)

\bibitem{Sch11}
U.~Schollw\"ock, Ann. Phys. (N.Y.) \textbf{326}, 96 (2011)

\bibitem{BRP+15}
J.~Braun, R.~Rausch, M.~Potthoff, J.~Minar, H.~Ebert, Phys. Rev. B \textbf{91},
  035119 (2015)

\bibitem{ATE+14}
H.~Aoki, N.~Tsuji, M.~Eckstein, M.~Kollar, T.~Oka, P.~Werner, Rev. Mod. Phys.
  \textbf{86}, 779 (2014)

\bibitem{GE16}
Y.~Ge, J.~Eisert, New J. Phys. \textbf{18}, 083026 (2016)

\bibitem{BPP13}
P.E. Bl\"ochl, T.~Pruschke, M.~Potthoff, Phys. Rev. B \textbf{88}, 205139
  (2013)

\end{thebibliography}
\end{document}